\documentclass[sort&compress]{jfm}

\usepackage{amsmath,amssymb}
\usepackage{graphicx}% Include figure files
\usepackage{dcolumn}% Align table columns on decimal point
\usepackage{bm}% bold math
\usepackage{hyperref}% add hypertext capabilities
\usepackage{siunitx}
\DeclareSIUnit\Molar{\textsc{M}}
\usepackage[noabbrev,nameinlink]{cleveref}
\usepackage[labelformat=simple]{subcaption}

\usepackage[normalem]{ulem}
\usepackage{comment}
\usepackage[usenames,dvipsnames]{color}
\usepackage{natbib}

\renewcommand{\vec}[1]{\bm{#1}}
\newcommand{\intd}{\mathop{}\!\mathrm{d}}
\newcommand{\abs}[1]{\left\lvert{#1}\right\rvert}

\newcommand{\bigO}[1]{\mathcal{O}\left({#1}\right)}
\newcommand{\ex}{\vec{e}_x}
\newcommand{\ey}{\vec{e}_y}
\newcommand{\ez}{\vec{e}_z}
\newcommand{\lap}[1]{\nabla^2{#1}}
\newcommand{\grad}[1]{\nabla{#1}}
\newcommand{\cross}{\times}
\newcommand{\ds}{\Delta\mathrm{s}}
\newcommand{\diff}[2]{\frac{\mathrm{d}#1}{\mathrm{d}#2}}
\newcommand{\Eh}{E_h}
\newcommand{\Sp}{Sp}
\newcommand{\frft}{\text{F-RFT}}
\newcommand{\wrft}{\text{W-RFT}}
\newcommand{\rss}{\text{RSS}}
\newcommand{\Spa}{$\Sp{}$--$\hat{a}$}

\shorttitle{Filament mechanics in a half-space via regularised Stokeslet segments}
\shortauthor{B. J. Walker, K. Ishimoto, H. Gad\^{e}lha and E. A. Gaffney}
\title{Filament mechanics in a half-space via regularised Stokeslet segments}

\author{B. J. Walker\aff{1}
  \corresp{\email{benjamin.walker@maths.ox.ac.uk}}, K. Ishimoto\aff{1,2}, H. Gad\^{e}lha\aff{3,4}
 \and E. A. Gaffney\aff{1}}

\affiliation{\aff{1}Wolfson Centre for Mathematical Biology, Mathematical Institute, University of Oxford, Oxford, OX2 6GG, UK
\aff{2}Graduate School of Mathematical Sciences, The University of Tokyo, Tokyo, 153-8914, Japan
\aff{3}Department of Mathematics, University of York, York YO10 5DD, UK
\aff{4}Department of Engineering Mathematics, University of Bristol, Bristol BS8 1UB, UK}

\begin{document}

\maketitle

\begin{abstract}
%!TEX root=../main.tex
We present a generalisation of efficient numerical frameworks for modelling
fluid-filament interactions via the discretisation of a recently-developed,
non-local integral equation formulation to incorporate regularised Stokeslets
with half-space boundary conditions, as motivated by the importance of
confining geometries in many applications. We proceed to utilise this
framework to examine the drag on slender inextensible filaments moving near a
boundary, firstly with a relatively-simple example, evaluating the accuracy of
resistive force theories near boundaries using regularised Stokeslet segments.
This highlights that resistive force theories do not accurately quantify
filament dynamics in a range of circumstances, even with analytical
corrections for the boundary. However, there is the notable and important
exception of movement in a plane parallel to the boundary, where accuracy is
maintained. In particular, this justifies the judicious use of resistive force
theories in examining the mechanics of filaments and monoflagellate
microswimmers with planar flagellar patterns moving parallel to boundaries. We
proceed to apply the numerical framework developed here to consider how
filament elastohydrodynamics can impact drag near a boundary, analysing in
detail the complex responses of a passive cantilevered filament to an
oscillatory flow. In particular, we document the emergence of an asymmetric
periodic beating in passive filaments in particular parameter regimes, which
are remarkably similar to the power and reverse strokes exhibited by motile
9+2 cilia. Furthermore, these changes in the morphology of the filament
beating, arising from the fluid-structure interactions, also induce a
significant increase in the hydrodynamic drag of the filament.
\end{abstract}

% Main text.
\section{Introduction}
\label{sec:intro}
%!TEX root=../main.tex
The mechanics of flexible filaments on the microscale underpin much of
biology, from the propulsive flagella of motile bacteria and spermatozoa to
nodal cilia, the latter hypothesised to be responsible for the breaking of
left-right symmetry in mammals \citep{Smith2019,Berg1973,Gray1928}.
Furthermore, the dynamics of elastic filaments are of intense interest in the
physics of microdevices and surface flows, including near-wall dynamics. For
instance, the consideration of soft deformable sensors has already motivated
extensive studies of attached filaments \citep{Guglielmini2012,Roper2006}, as
has the characterisation of attached filament forces for understanding the
drag induced by slender appendages
\citep{Curtis2012,Simons2014,Pozrikidis2011}. Such appendages range from the
primary cilium to carbon nanotube mats, with an extensive review of the field
presented by \citet{DuRoure2019}, which notes that both theoretical and
numerical developments are very much still required in this field. Indeed,
with advances in microscopy enabling ever more detailed quantification of
kinematics, often with confining geometry such as a cover slip or substrate,
the development of validated, and ideally simple, methodologies would be
beneficial in estimating mechanics from kinematics.

Furthermore, the complex mechanics of fluid-structure interaction is an
important problem and has been well studied, as illustrated by the slender
body theory of \citet{Tornberg2004} and \citet{Liu2018}. Numerical simulations
that attempt to move past slender body theory are frequently plagued by
extensive numerical stiffness, such as the regularised Stokeslet simulations
of \citet{Ishimoto2018a} and \citet{Olson2013}. The recent advance of
\citet{Moreau2018} and its subsequent extension by \citet{Hall-McNair2019}
have sought to address this stiffness, with their methodologies significantly
reducing the computational cost associated with filament-fluid interactions
via the use of integrated force and moment balance equations, but have not
considered even the simplest confined geometry that is an infinite planar
wall. Hence our first objective will be to generalise these improved
frameworks to dynamics in a half-space bounded by a wall, adapting the recent
regularised Stokeslet segment approach of \citet{Cortez2018} for drag
calculations and elastohydrodynamics. We will validate in detail the
computation of drag from kinematic data against the earlier work of
\citet{Ramia1993}, additionally validating our proposed elastohydrodynamic
framework against the gold-standard boundary element method of
\citet{Pozrikidis2010,Pozrikidis2011}.

These validations demonstrate the high accuracy of our approach in capturing
the mechanics of filaments, and thus it is of extensive use for understanding
cellular swimming. For instance, resistive force theories are very popular due
to their ease of use
\citep{Moreau2018,Lauga2006,Utada2014,Sznitman2010,Schulman2014,Gadelha2010},
and they have been shown to be of reasonable accuracy in free space for
small-bodied cells such as spermatozoa \citep{Johnson1979}. With freedom to
choose the resistive coefficients, such local drag theories have been shown to
perform very well even near a boundary \citep{Friedrich2010}, as supported by
the boundary element method validation studies for straight rods of
\citet{Ramia1993}. This suggests that the refinements of \cite{Katz1975} and
\citet{Brenner1962} to free-space resistive force theories, valid for straight
filament motion parallel or perpendicular to a planar wall, may be useful for
the analysis of microscopy data, and would be very popular if demonstrated to
be accurate. In particular, in the common circumstance where subject cells are
imaged swimming parallel to a coverslip, for example
\citet{Friedrich2010,Riedel-Kruse2007}, and thus have unchanging boundary
separation, we hypothesise that these refined resistive force theories may be
sufficient to accurately capture the hydrodynamic drag on a curved filament.
Hence, as our first application, we use the regularised Stokeslet segment
framework of \citet{Cortez2018} to test this hypothesis in exemplar problems,
both when a filament is moving parallel to a boundary and when it is attached
to the surface, subsequently moving extensively perpendicular to the wall.

Further, the presented elastohydrodynamic framework is sufficiently flexible
to enable sophisticated considerations of fluid-structure interactions,
inherited from the approach of \citet{Moreau2018}. Hence, as an application of
non-local elastohydrodynamics in a half-space, we proceed to demonstrate the
methodology's ease of use in the context of such complex physics by examining
the mechanics of a flexible inextensible adhered filament in an oscillatory
flow, generalising \citeauthor{Pozrikidis2011}' study of carbon nanomat
surface drag to time dependent flows \citep{Pozrikidis2011}. Such simulations
are clear precursors for the application of the framework to simulations of
oscillating systems in physiological and soft matter modelling, such as
primary and motile ciliary systems respectively in the kidneys and lungs, in
addition to numerical studies of active surfaces, for example
\citet{Shum2013,Balazs2014}.
\section{Methods}
\label{sec:methods}
%!TEX root=../main.tex
\subsection{Piecewise-linear filaments}\label{sec:methods:filament_disc}
\begin{figure}
	\centering
	\vspace{1em}
	\includegraphics[width=0.7\textwidth]{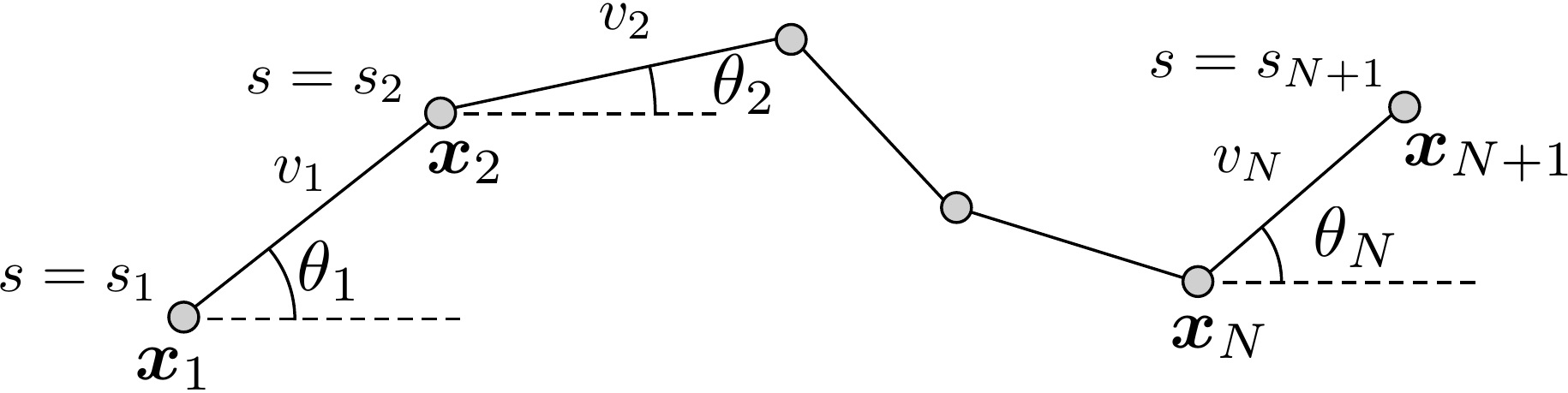}
	\caption{The two-dimensional piecewise-linear filament. We consider an
	inextensible filament composed of $N$ straight segments each of length
	$v_i$, connected at material points $\vec{x}_i$ with each segment making
	angle $\theta_i$ with the global $\ex$ axis. The endpoints $\vec{x}_1$ and
	$\vec{x}_{N+1}$ shall be referred to as the base and tip respectively, and
	we note that, given segment lengths $v_i$, the entire filament may be
	described by the location of the base and the angles $\theta_i$.
	\label{fig:methods:discretised_filament}}
\end{figure}

% Basics of discretisation necessary for both parts, including tangent angle
% here. Note we solve for a filament made of straight links, rather than
% approximate the continuum directly. 
We consider a planar slender inextensible filament represented by
$N$ piecewise-linear segments of constant length and described by arclength
parameter $s\in[0,L]$, where $L$ is the filament length.% For a slender filament described at time $t$ by a smooth plane curve
% $\vec{x}(s,t)$ with arclength parameter $s\in[0,L]$, we consider a
% discretisation of $\vec{x}(s,t)$ into $N$ piecewise-linear segments of
% constant length. 
The segment endpoints are taken to correspond to material points, with their
positions being denoted $\vec{x}_1,\ldots,\vec{x}_{N+1}$ and where $\vec{x}_i$
and $\vec{x}_{i+1}$ correspond to the $i$\textsuperscript{th} segment. We will
refer to $\vec{x}_1$ as the base of the filament and correspondingly
$\vec{x}_{N+1}$ as the tip, and denote the constant arclength associated with
the material point $\vec{x}_i$ as $s_i$. As the filament is assumed to be
planar, without loss of generality we assume that it lies in a plane spanned
by orthogonal unit vectors $\ex$ and $\ey$, with $\ez$ completing the
orthonormal right-handed triad $\ex\ey\ez$, and we may write $\vec{x}_i=x_i\ex
+ y_i\ey$. Following \citet{Moreau2018} we note that, once the segment lengths $v_1,\ldots,v_N$ are prescribed, the filament may be
completely described by the $N+2$ scalars
$x_1,y_1,\theta_1,\ldots,\theta_N$, where $\theta_i$ is defined as the angle
between the $i$\textsuperscript{th} segment and the unit vector $\ex$ as
shown in \cref{fig:methods:discretised_filament}. Explicitly, from
these $N+2$ variables we may recover the filament endpoints as
\begin{equation}
	\vec{x}_j = \vec{x}_1 + \sum\limits_{i=1}^{j-1} \left(\cos{\theta_i}\ex + \sin{\theta_i}\ey\right)v_i\,,
\end{equation}
and, using dots to denote derivatives with respect to time, the velocity of
each material point is given by
\begin{equation}
	\dot{\vec{x}}_j = \dot{\vec{x}}_1 + \sum\limits_{i=1}^{j-1} \left(-\sin{\theta_i}\ex + \cos{\theta_i}\ey\right)\dot{\theta}_iv_i\,,
\end{equation}
again expressible using only the $N+2$ variables due to the imposed
geometrical constraints.

\subsection{Force distributions via regularised Stokeslet segments}
\label{sec:methods:force_given_kinematics}
% Recapitulate relevant sections of Cortez, include in Supplementary Material
% the evaluation of the $1/R^5$ integral.
To describe the low-Reynolds number fluid dynamics pertinent to the filament
we utilise the recent work of \citet{Cortez2018}, namely the method of
regularised Stokeslet segments (\rss{}). Here we briefly recapitulate the
formulation of this methodology as presented by \citeauthor{Cortez2018},
relating the force density applied on the fluid by the filament to the fluid
velocity via non-local hydrodynamics.

As in the popular method of regularised Stokeslets \citep{Cortez2001}, we
begin by considering solutions of the three-dimensional smoothly-forced Stokes
equations, which may be stated for viscosity $\mu$ and force $\vec{f}$ applied
on the fluid at the origin as
\begin{equation}\label{eq:methods:smoothly_forced_stokes}
	\mu\lap{\vec{u}} = \grad{p} - \vec{f}\phi_{\epsilon}\,, \qquad p,\vec{u}\rightarrow 0 \, \text{ as } \abs{\vec{x}}\rightarrow\infty\,,
\end{equation}
where $\vec{u}$ is the fluid velocity, $p$ is the pressure and
$\phi_{\epsilon}$ is a smooth approximation to the Dirac delta distribution
dependent on the small parameter $\epsilon$. Following section 2 of \citet{Cortez2018} we take
\begin{equation}
	\phi_{\epsilon}(\vec{x}) = \frac{15\epsilon^4}{8\pi\left(\abs{\vec{x}}^2 + \epsilon^2\right)^{\frac{7}{2}}}\,,
\end{equation}
for which the solution to \cref{eq:methods:smoothly_forced_stokes} is known and given by
\begin{equation}
	8\pi\mu\vec{u}(\vec{x}) = \left[\left(\frac{1}{R} + \frac{\epsilon^2}{R^3}\right)I + \frac{\vec{x}\vec{x}^T}{R^3}\right]\vec{f}\,, \quad R(\vec{x}) = \sqrt{\abs{\vec{x}}^2 + \epsilon^2}\,.
\end{equation}
By linearity of the Stokes equations the velocity at a point $\hat{\vec{x}}$
due to a distribution of regularised Stokeslets along the filament with force
density $\vec{f}(s)$ is therefore given by
\begin{equation}
	8\pi\mu\vec{u}(\hat{\vec{x}}) = \int\limits_0^L \left[\left(\frac{1}{R(\vec{r})} + \frac{\epsilon^2}{R(\vec{r})^3}\right)I + \frac{\vec{r}\vec{r}^T}{R(\vec{r})^3}\right]\vec{f}(s)\intd{s}\,, \quad \vec{r}(s) = \hat{\vec{x}}-\vec{x}(s)\,.
\end{equation}
Whilst the original method of regularised Stokeslets would entail
approximating the force distribution by a finite number of smoothed point
forces, the method of regularised Stokeslet segments instead considers a
linear distribution of forces along the straight segments of the discretised
filament. With the discretisation of the force distribution as a continuous
piecewise-linear function  along each of the segments, the fluid velocity is
instead given by a sum of integrals over each of the segments, where each
individual integral may be analytically evaluated. Parameterising the
$j$\textsuperscript{th} segment by $\alpha\in[0,1]$ so that $\vec{x} =
\vec{x}_j - \alpha\vec{v}$ for $\vec{v} = \vec{x}_j - \vec{x}_{j+1}$, and
additionally writing the force density as $\vec{f} = \vec{f}_{j} +
\alpha(\vec{f}_{j+1}-\vec{f}_j)$ for force densities $\vec{f}_j,\vec{f}_{j+1}$
at $\vec{x}_j,\vec{x}_{j+1}$ respectively, the integral along the $j$\textsuperscript{th} segment is given by
\begin{equation}\label{eq:methods:integral_along_segment}
	v_j\int\limits_0^1 \left[\left(\frac{1}{R} + \frac{\epsilon^2}{R^3}\right)I + \frac{(\hat{\vec{x}}-\vec{x}_j+\alpha\vec{v})(\hat{\vec{x}}-\vec{x}_j+\alpha\vec{v})^T}{R^3}\right][\vec{f}_{j} +
\alpha(\vec{f}_{j+1}-\vec{f}_j)]\intd{\alpha}\,,
\end{equation}

where we have identified $\abs{\vec{v}} = v_j$ and
$R=\sqrt{\abs{\hat{\vec{x}}-\vec{x}_j+\alpha\vec{v}}^2+\epsilon^2}$. As noted
by \citet[sec. 2.2]{Cortez2018} this may be written as a sum of integrals of
the form
\begin{equation}
	T_{m,p} = \int\limits_0^1 \alpha^m R^p\intd{\alpha}
\end{equation}
for $(m,p)\in\{(0,-1),(1,-1),(0,-3),(1,-3),(2,-3),(3,-3)\}$, each of which is
implicitly dependent on $j$ and may be evaluated by explicit computation of
$T_{0,-1}$ and $T_{0,-3}$ and subsequent application of the recurrence
relation
\begin{equation}\label{eq:methods:recurrence}
	v_j^2 T_{m+1,p} = -\frac{m}{(p+2)}T_{m-1,p+2} - (\hat{\vec{x}}-\vec{x}_j)\cdot\vec{v}T_{m,p} + \frac{1}{p+2}\alpha^m\left.R^{p+2}\right\rvert_0^1\,, \quad p\neq-2\,,
\end{equation}
which follows from the definition of $T_{m,p}$ and consideration of $\diff{}{\alpha}\left(\alpha^mR^p\right)$.

\begin{figure}
\centering
	\begin{subfigure}[c]{0.45\textwidth}
		\centering
		\includegraphics[width=0.8\textwidth]{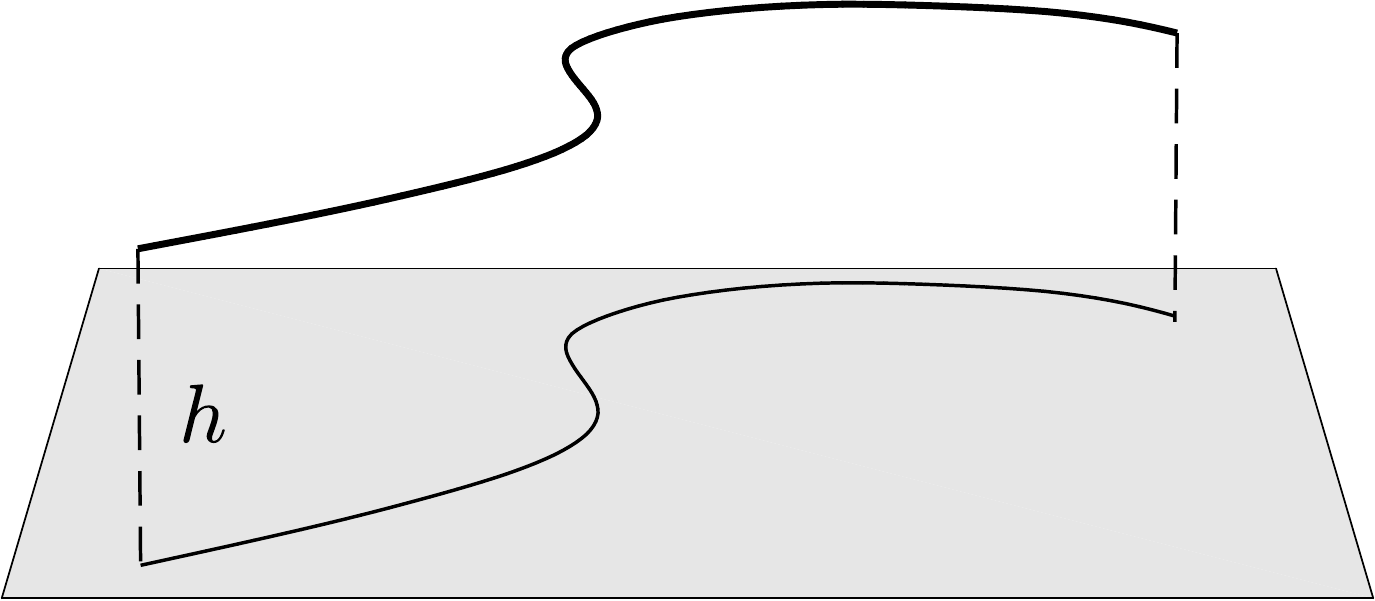}
		\caption{\label{fig:methods:wall_configurations:parallel}}
	\end{subfigure}
	\begin{subfigure}[c]{0.45\textwidth}
		\centering
		\vspace{1.01em}
		\includegraphics[width=0.8\textwidth]{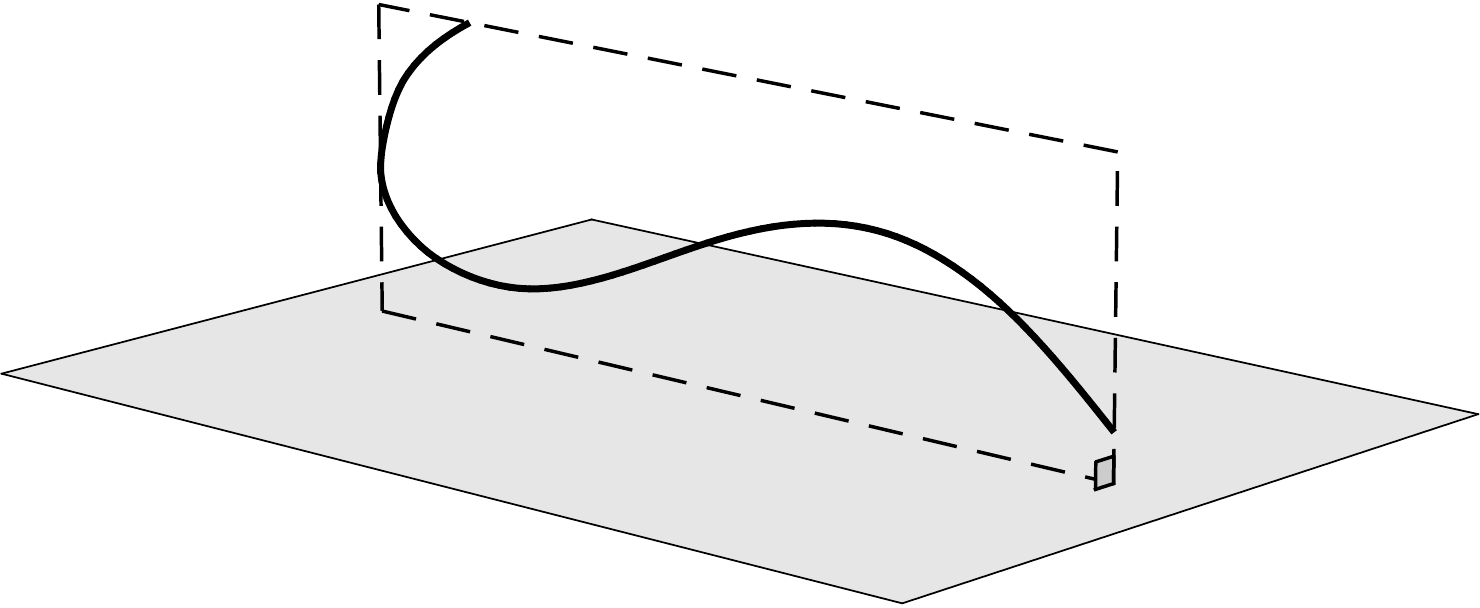}
		\caption{\label{fig:methods:wall_configurations:perpendicular}}
	\end{subfigure}
	\caption{Schematic configurations of planar filaments near boundaries.
	\subref{fig:methods:wall_configurations:parallel} A filament moving
	parallel to the infinite planar boundary, at constant separation $h$. The
	projection of the filament onto the boundary is shown as a thin black
	line. \subref{fig:methods:wall_configurations:perpendicular} A filament
	contained in a plane perpendicular to the boundary, with the plane of
	motion shown dashed.
	\label{fig:methods:wall_configurations}}
\end{figure}

The simple extension of this methodology to a half-space bounded by a no-slip
planar wall via the image system of \citet{Ainley2008} (with the typographical
error corrected \citep{Smith2009d}) is presented in detail by \citet[sec.
3.3]{Cortez2018}. Exemplified in \cref{fig:methods:wall_configurations}, we
will consider two configurations of half space, one in which the infinite
planar boundary is parallel to the plane containing the filament, given by
$z=0$, and another where the boundary is situated perpendicular to this plane
and given by $y=0$. In both cases the hydrodynamics may be described by the
same regularised singularity representation, each requiring computation of
$T_{m,p}$ for the additional values
$(m,p)\in\{(0,-5),(1,-5),(2,-5),(3,-5),(4,-5),(5,-5)\}$ in order to include
the additional necessary regularised singularities. Omitted from the previous
work of Cortez, we compute
\begin{equation}
 	T_{0,-5} = \left.\frac{(B+\alpha C^2)}{3R^3} \frac{\left(-B^2 + C^2\left[3A^2+4\alpha B + 2\alpha^2C^2 + 3\epsilon^2\right]\right)}{\left(B^2 - C^2\left[A^2+\epsilon^2\right]\right)^2}\right\rvert_0^1\,,
\end{equation}
where $A=\abs{\hat{\vec{x}}-\vec{x}_j}$,
$B=(\hat{\vec{x}}-\vec{x}_j)\cdot\vec{v}$, and $C = \abs{\vec{v}}=v_j$. The
remaining values of $T_{m,p}$ may be computed via the recurrence relation
\cref{eq:methods:recurrence}. Simple computation of the coefficients of the
$T_{m,p}$ results in a matricial form of the velocity contribution at
$\hat{\vec{x}}$ from the $j$\textsuperscript{th} segment as a linear operator
acting on $\vec{f}_j$ and $\vec{f}_{j+1}$, the details of which are cumbersome
and given in the Supplementary Material. The overall velocity at the
evaluation point $\hat{\vec{x}}$ is then given by the sum of these
contributions, resulting in a matricial equation of the form
\begin{equation}
	\vec{u}(\hat{\vec{x}}) = M(\hat{\vec{x}}) \vec{F}\,,
\end{equation}
where $\vec{F}=(f_{1,x},f_{1,y},\ldots,f_{N+1,x},f_{N+1,y})^T$ herein denotes
the composite vector of force densities applied on the fluid at the segment
endpoints $\vec{x}_j$ in the directions $\ex$ and $\ey$, where we write
$\vec{f}(\vec{x}_j)=f_{j,x}\ex + f_{j,y}\ey$. Taking the evaluation point
$\hat{\vec{x}}$ as each $\vec{x}_i$ in turn yields a square system of linear
equations in $2(N+1)$ variables, relating the velocities $\vec{u}(\vec{x}_i)$
to the force distribution on the filament via non-local hydrodynamics.
Application of the no-slip condition at the segment endpoints then gives a
relation between the filament velocities and the forces applied on the fluid
by the filament, which we write in brief as
\begin{equation}
	\dot{\vec{X}} = A \vec{F}\,,
\end{equation}
for the square matrix $A$ composed row-wise of blocks $M(\vec{x}_i)$ for
$i=1,\ldots,N+1$, and where
$\dot{\vec{X}}=(\dot{x}_1,\dot{y}_1,\ldots,\dot{x}_{N+1},\dot{y}_{N+1})^T$.
Given kinematic data, this linear system may be readily solved to give the
force densities applied on the fluid by the filament.

\subsection{Extension to efficient solution of elastohydrodynamic equations}\label{sec:methods:elastohydrodynamic}
% Formulate the problem of solving the full elastohydrodynamics.
We now proceed to combine the work of \citet{Moreau2018} and
\citet{Cortez2018} to give an efficient scheme for solving non-local planar
elastohydrodynamics, similar in concept to the piecewise-constant force
density approach of \citet{Hall-McNair2019} but with continuous
piecewise-linear force discretisation and the additional inclusion of an
infinite planar boundary. As formulated by \citet{Moreau2018}, we integrate
the pointwise conditions of force and moment balance on the filament, given by
\begin{align}
	\vec{n}_s - \vec{f} &= \vec{0}\,,\\
	\vec{m}_s + \vec{x}_s \cross \vec{n} &= \vec{0}
\end{align}
for contact force and couple denoted $\vec{n},\vec{m}$ respectively and where
a subscript of $s$ denotes differentiation with respect to arclength, noting
that we have assumed slenderness of the filament. This yields the integrated
equations
\begin{align}
	- \sum\limits_{j=1}^{N} \int\limits_{s_j}^{s_{j+1}} \vec{f}(s)\intd{s} &= \vec{n}(0)\,, \\ 
	- \sum\limits_{j=i}^{N}\int\limits_{s_j}^{s_{j+1}} (\vec{x}(s)-\vec{x}_i)\cross\vec{f}(s)\intd{s} &= \vec{m}(s_i)\,, \quad i = 1,\ldots,N\,.
\end{align}
Here we have assumed conditions of zero contact force and couple at the tip of
the filament, i.e. $\vec{n}(L)=\vec{m}(L)=\vec{0}$, retaining generality at
the base for the time being. With the assumption of piecewise-linear
distributions of force density $\vec{f}$ over each segment, we may write these
integrals as linear operators on $\vec{F}$, yielding the system
\begin{equation}
	-B\vec{F} = \vec{R}\,,
\end{equation}
where $\vec{R} =
(\vec{n}(s_1)\cdot\ex,\vec{n}(s_1)\cdot\ey,m(s_1),\ldots,m(s_N))^T$ and we are
writing $\vec{m}(s_i)=m(s_i)\ez$ by planarity of the filament. The matrix $B$
is of dimension $(N+2) \cross 2(N+1)$, which under the assumption of
equispaced segment endpoints has rows $B_k$ given by
\begin{align}
	B_1 & = \frac{\ds}{2}\left[1,0,2,0,2,\ldots,2,0,1,0\right]\,, \\
	B_2 & = \frac{\ds}{2}\left[0,1,0,2,0,2,\ldots,2,0,1\right]\,,
\end{align}
\begin{multline}
	B_{i+2} = \ds\big[\underbrace{0,\ldots,0}_{2(i-1)},\, -\frac{\ds}{6}\sin{\theta_i},\, \frac{\ds}{6}\cos{\theta_i},\\
	\underbrace{\ldots,-y_j+y_i-\frac{\ds}{2}\sin{\theta_j},\, x_j-x_i+\frac{\ds}{2}\cos{\theta_j},\ldots}_{j=i+1,\ldots,N}\\
	-\frac{1}{2}(y_N-y_i)-\frac{\ds}{3}\sin{\theta_N},\, \frac{1}{2}(x_N-x_i) + \frac{\ds}{3}\cos{\theta_N}\big]\,,
\end{multline}
where $i=1,\ldots,N$ and $\ds=L/N$ is the length of each segment. Thus we may
write the coupled elastohydrodynamic problem as
\begin{equation}
	-BA^{-1}\dot{\vec{X}}=\vec{R}\,,
\end{equation}
where $B$ represents integration over segments and $A$ encodes the
relationship between the forces on the surrounding fluid and the velocities of
material points on the filament, here non-local and assumed to be invertible.

As noted in \cref{sec:methods:filament_disc}, the material velocities
$\dot{\vec{x}}_i$ may be expressed in terms of the time derivatives of the
base point $\vec{x}_1$ and the segment angles $\theta_1,\ldots,\theta_N$. Defining $Q$ to be the $2(N+1)\cross(N+2)$ matrix such that 
\begin{equation}
	Q\dot{\vec{\theta}}=\dot{\vec{X}}\,,
\end{equation}
where $\vec{\theta}=\left(x_1,y_1,\theta_1,\ldots,\theta_N\right)^T$, we have $Q$
given explicitly by
\begin{equation}
	Q = \left[
	\begin{array}{cc|ccc}
	1 & 0 & &&\\
	\vdots & \vdots && Q_1&\\
	1 & 0 & && \\
	\hline
	0 & 1 & &&\\
	\vdots & \vdots && Q_2&\\
	0 & 1 & &&
	\end{array}\right]_P\,,
\end{equation}
where the subscript $P$ denotes that the $i$\textsuperscript{th} row of $Q$ is
to be permuted to
\begin{equation}
	P(i)=\left\{\begin{array}{rl}
	2(i-1)+1\,, & i = 1,\ldots,N+1\,,\\
	2(i-N-1)\,, & i = N+2,\ldots,2N+2\,.
\end{array}\right.
\end{equation}
This permutation of $Q$ allows us to define the blocks $Q_1$ and $Q_2$ as
being strictly lower triangular matrices of dimension $(N+1)\cross N$ with
entries
\begin{align}
	Q_1^{i,j} &=\ds\left\{\begin{array}{rl}
		-\sin{\theta_j}\,, & j<i\,,\\
		0\,, & j\geq i\,,
	\end{array}\right. \\
	Q_2^{i,j} &=\ds\left\{\begin{array}{rl}
		+\cos{\theta_j}\,, & j<i\,,\\
		0\,, & j\geq i\,.
	\end{array}\right.
\end{align}
Given the matrix $Q$ we may rewrite the full coupled elastohydrodynamic
problem simply as
\begin{equation}
	-BA^{-1}Q\dot{\vec{\theta}}=\vec{R}\,,
\end{equation}
with $BA^{-1}Q$ being a square matrix of dimension $(N+2)\cross(N+2)$.
Finally, and here for example assuming force and moment-free conditions at the
filament base, the standard linear constitutive relation
$m(s_i)=EI\theta_s(s_i)\approx EI(\theta_i-\theta_{i-1})/\ds$ for bending
stiffness $EI$, valid for $i=2,\ldots,N$, gives $\vec{R}$ explicitly in terms
of $\vec{\theta}$, yielding a linear system of low dimension that may be
readily solved for $\dot{\vec{\theta}}$, with the temporal dynamics then
computable via existing ODE methods. As noted by \citeauthor{Moreau2018}, this
formulation is readily extensible to the imposition of a variety of boundary
conditions, in particular that of a cantilevered filament base, achieved by
replacing the overall zero-force and zero-moment equations with
$\dot{x}_1=\dot{y}_1=\dot{\theta}_1=0$.

We non-dimensionalise this system by scaling spatial coordinates with filament
length $L$, forces with $EI/L^2$, and time with some characteristic timescale
$T$, yielding the non-dimensional system
\begin{equation}
	-\Eh \hat{B}\hat{A}^{-1}\hat{Q}\dot{\hat{\vec{\theta}}} = \hat{\vec{R}}\,, \quad E_h = \frac{8\pi\mu L^4}{EI\cdot T}
\end{equation}
for elastohydrodynamic number $E_h$, where the notation $\hat{\cdot}$ denotes
non-dimensional quantities. The dimensional and non-dimensional quantities are
related by
\begin{equation}
	B = L^2 \hat{B}\,, \quad A = \frac{1}{8\pi\mu}\hat{A}\,, \quad Q\dot{\vec{\theta}} = \frac{L}{T}\hat{Q}\dot{\hat{\vec{\theta}}}\,, \quad \vec{R} = \frac{EI}{L}\hat{\vec{R}}\,,
\end{equation}
where we have multiplied the two force balance equations by $\ds=L/N$ and
absorbed the dimensional scalings of $x_1,y_1$ into $Q$ for convenience, with
$\hat{\vec{\theta}}=(\hat{x}_1,\hat{y}_1,\theta_1,\ldots,\theta_N)^T$. In
order to pose the non-dimensional equations in terms of the commonly-used
sperm number $\Sp$ \citep{Moreau2018,Delmotte2015,Ishimoto2018a}, we proceed
following \citet{Ishimoto2018a} to define
\begin{equation}
 	\Sp^4 = \frac{\xi L^4}{EI\cdot T}\,, \quad \xi=\frac{4\pi\mu}{\log(2L/\epsilon)}
\end{equation}
for resistive force coefficient $\xi$ and $L/\epsilon$ = $10^{3}$, giving the
approximate relation $\Eh \approx 15.2 \cdot\Sp^4$.

\subsection{Implementation, verification and parameter choice}
% Verify far-field implementation and note that boundary condition is
% satisfied. Additional verification? Verify forward and reverse problem.
% Briefly present dependence on epsilon and N.
% Compare against RFT for straight filament case!
Both the calculation of force densities from kinematic data and the solution
of full elastohydrodynamics were implemented in MATLAB, the latter utilising
the inbuilt stiff ODE solver \texttt{ode15s} \citep{Shampine1997}. Prior to
the recent work of \citet{Hall-McNair2019} the solution of non-local
elastohydrodynamics has required significant computational work and minimal
timestep for the solution of this stiff problem
\citep{Olson2013,Ishimoto2018a}, and we replicate in our implementation the
low computational cost associated with the integrated elasticity equations of
\citet{Moreau2018}. In particular, typical simulations of a cantilevered
filament in background flow, explored in detail in
\cref{sec:results:oscillatory_morphological,sec:results:oscillatory_drag},
have a typical runtime of $10\si{\s}$, where $N=40$ and we simulate over 10
periods of oscillation of the background flow on modest hardware
(Intel\textsuperscript{\textregistered} Core\texttrademark\ i7-6920HQ CPU).

Verification was first performed on the reduction to an unbounded domain, with
the full elastohydrodynamics being verified by comparison with the resistive
force theory results of \citet{Moreau2018} for filament relaxation. Further
drag calculations were compared against the work of \citet{Cortez2018} for the
case of uniform motion. The regularised image system in a half-space was then
verified by explicit evaluation of the fluid velocity on the no-slip boundary,
with the numerical result being zero to machine precision, and further by
noting that far-field results were seen to converge to those of the free-space
system. Further qualitative checks were performed, an example being the
successful reproduction of the intuitive result that relaxation timescales
increase for filaments with reduced separation from the no-slip stationary
boundary, along with numerical comparison against the previous works of
\citeauthor{Pozrikidis2011} regarding filaments in steady shear flow
\citep{Pozrikidis2010,Pozrikidis2011}. Results of additional verification
against the boundary element method of \citet{Ramia1993} are shown in
\cref{fig:results:drag_compare_uniform_filament:close}. Sufficient accuracy is
typically achieved with $N=20$ segments, similar to the discretisation used by
\citet{Cortez2018}, though we typically take $N=40$ to enable the capturing of
high-curvature filaments.

As noted by \citet{Cortez2018}, the use of regularised Stokeslet segments
allows the regularisation parameter to represent the radius of the filament
being modelled, verified here by comparison with the boundary element method
of \citet{Ramia1993} in \cref{fig:results:drag_compare_uniform_filament},
subject to the phenomenological condition that $\epsilon$ be less than the
length of the segments, which appears necessary for convergence. We will take
$\epsilon/L = 10^{-3}$ unless otherwise stated, with typical slender filament
aspect ratios yielding $\epsilon/L$ in the range $(10^{-2},10^{-3})$
\citep{Ishimoto2016a,Yonekura2003}. Whilst the results that follow naturally
depend on the size the of the regularisation parameter, it being used as a
proxy for the filament radius, this dependence appears to be predominantly
quantitative, with the same qualitative conclusions holding across the range
of physically-relevant values of $\epsilon/L$.

\subsection{Endpoint effects}\label{sec:methods:endpoint_effects}
Inherent to the method of regularised Stokeslet segments as presented here is
the presence of apparent large variations in computed force densities at the
tips of filaments. This was initially observed by \citet[sec. 3.1, fig.
4]{Cortez2018}, who remarked that these endpoint effects may be reduced when
$\epsilon/L$ is small. Indeed, for $\epsilon/L$ in our range of physical
interest these endpoint effects contribute minimally to overall drag
calculations, in particular having little effect on the computed total drag
exerted on a filament and the computed force densities away from the filament
tip, whose presentation we focus on herein. Exploration of the effects of
non-uniform segment lengths yields the observation that these oscillatory
endpoint effects remain limited to approximately the 3 segments proximal to
the ends of the filament, thus the integral contribution of these oscillations
may be reduced by the clustering of segments near the filament tips. For our
typical parameters of $N=40$ and $\epsilon/L=10^{-3}$ we observe a difference
in integrated drag along of filaments of less than 2\% between linear,
quadratic and Chebyshev segment endpoints, hence we proceed with calculations
utilising segments of uniform length.% We do not
% attempt to report on the detailed computation of force densities at the
% filament tips, with such calculations being of little physical relevance
% without careful and extended consideration of filament endpoint geometry.

\subsection{Boundary-corrected resistive force theory}
The leading-order approximation of slender body hydrodynamics that is
resistive force theory (RFT) was first introduced by Gray and Hancock
\citep{Hancock1953,Gray1955}, relating the local drag on a body to its local
tangential and normal velocities $u_t, u_n$ via
\begin{equation}
	-C_t u_t = f_t\,, \quad  -C_n u_n = f_n\,.
\end{equation}
Here $f_t,f_n$ are the local tangential and normal components of force applied
on the fluid, with the constant resistive coefficients $C_t,C_n$ typically
being functions of filament aspect ratio. As RFT relates forces only to local
velocities, calculations using this simple local drag theory are not able to
account for the presence of a boundary. \citet{Brenner1962} and
\citet{Katz1975} posed corrections to resistive force theory for straight
filaments in asymptotic regimes far-from or near-to an infinite no-slip
boundary, with the latter having been verified against high-accuracy boundary
element methods by \citet{Ramia1993}. In the case of filament motion confined
to a plane parallel to the boundary, as exemplified in
\cref{fig:methods:wall_configurations:parallel}, by this wall-corrected
resistive force theory (\wrft) a straight filament parallel to the boundary
has resistive coefficients given by
\begin{align}
	C_t &= \frac{2\pi\mu}{\log{\left(\frac{2}{\epsilon}\right)} -0.807 - \frac{3L}{8h}}\,, && 
	C_n = \frac{4\pi\mu}{\log{\left(\frac{2}{\epsilon}\right)} +0.193 - \frac{3L}{4h}}  &&
	\text{if }L \ll h \label{eq:methods:rft_corrected_para_far}\,,\\
	C_t &= \frac{2\pi\mu}{\log{\left(\frac{2h}{\epsilon}\right)}}\,, &&
	C_n = \frac{4\pi\mu}{\log{\left(\frac{2h}{\epsilon}\right)}} &&
	\text{if }L \gg h \label{eq:methods:rft_corrected_para_close}\,,
\end{align}
where $h$ is the boundary separation of the filament and $\epsilon$ has been
taken as the filament radius. Similarly, and as in
\cref{fig:methods:wall_configurations:perpendicular}, when filament motion is
in a plane perpendicular to the boundary the coefficients are given by
\begin{align}
	C_t &= \frac{2\pi\mu}{\log{\left(\frac{2}{\epsilon}\right)} -0.807 - \frac{3L}{8h}}\,, && 
	C_n = \frac{4\pi\mu}{\log{\left(\frac{2}{\epsilon}\right)} +0.193 - \frac{3L}{2h}}  &&
	\text{if }L \ll h \label{eq:methods:rft_corrected_perp_far}\,,\\
	C_t &= \frac{2\pi\mu}{\log{\left(\frac{2h}{\epsilon}\right)}}\,, &&
	C_n = \frac{4\pi\mu}{\log{\left(\frac{2h}{\epsilon}\right)}-1} &&
	\text{if }L \gg h \label{eq:methods:rft_corrected_perp_close}\,,
\end{align}
where all coefficients are as summarised by \citet{Brennen1977} and
\cref{eq:methods:rft_corrected_perp_close,eq:methods:rft_corrected_para_close}
additionally require $\epsilon\ll h$. In free-space we will adopt the
resistive force coefficients defined by the large-$h$ limit of
\cref{eq:methods:rft_corrected_para_far,eq:methods:rft_corrected_perp_far},
and refer to this simpler theory as free-space resistive force theory (\frft).
As given by \citeauthor{Brenner1962} and \citeauthor{Katz1975}, the leading
order boundary corrections to $C_t$ and $C_n$ are $\bigO{L/h}^3$ when $L \ll
h$, and $\bigO{h/L}$ when $L \gg h$, though the overall error in the resistive
force approximation remains logarithmic in the filament aspect ratio.
\section{Results and Applications}
\label{sec:results}
%!TEX root=../main.tex
\subsection{Evaluation of wall-corrected resistive force theory for straight filaments}
\label{sec:results:straight}
\begin{figure}
\centering
	\begin{subfigure}[c]{0.45\textwidth}
		\centering
		\includegraphics[width=\textwidth]{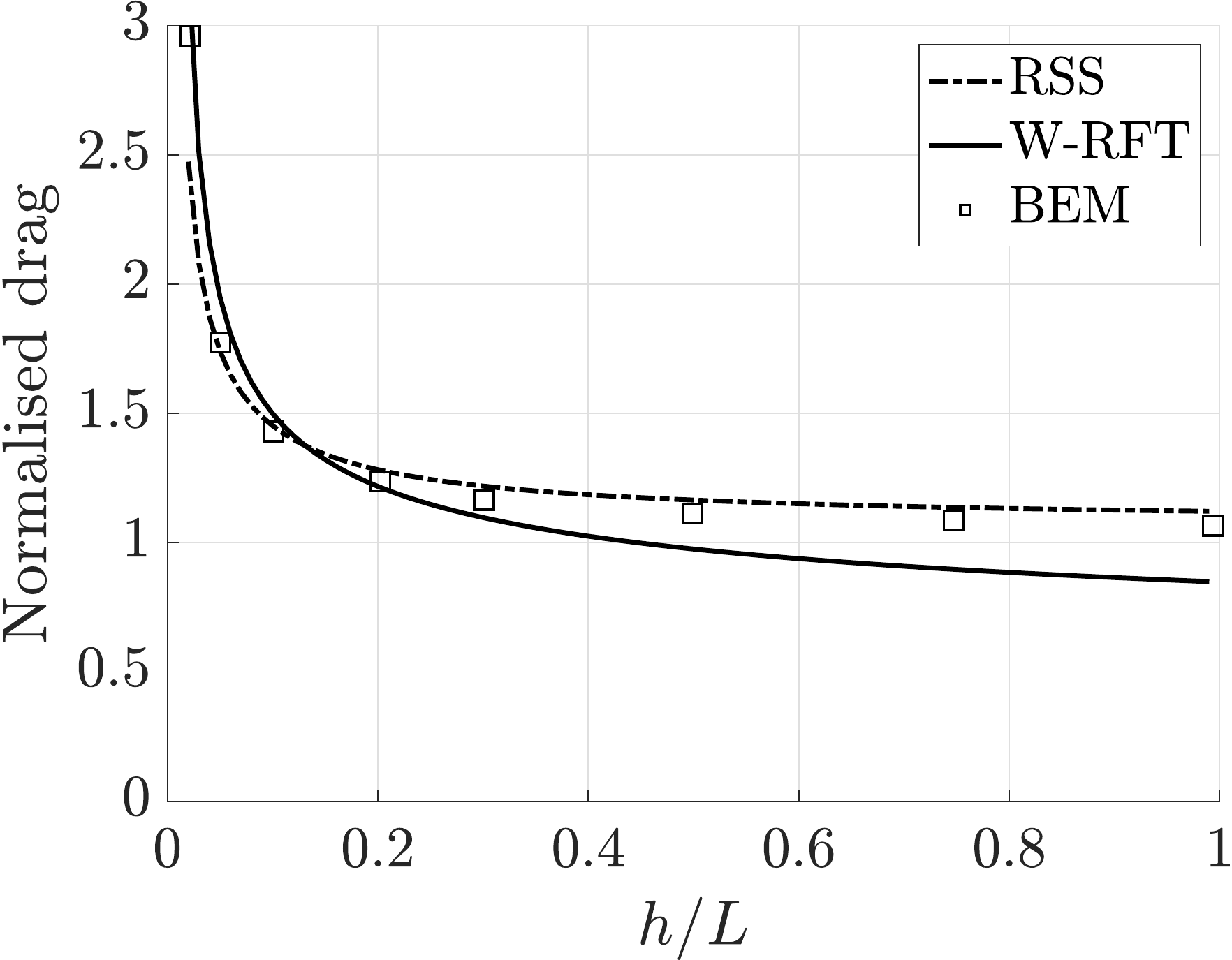}
		\caption{\label{fig:results:drag_compare_uniform_filament:close}}
	\end{subfigure}
	\begin{subfigure}[c]{0.45\textwidth}
		\centering
		\includegraphics[width=\textwidth]{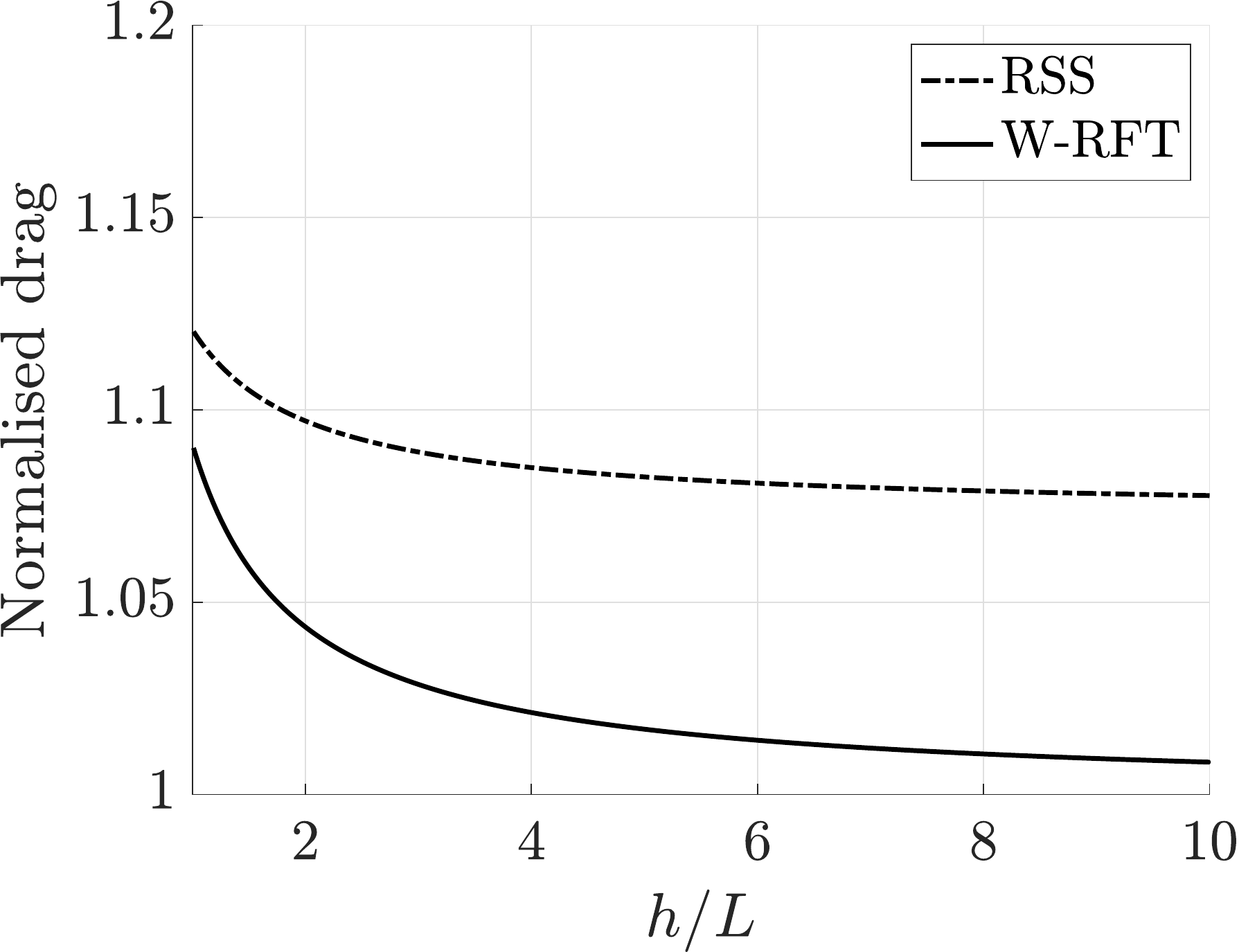}
		\caption{\label{fig:results:drag_compare_uniform_filament:far}}
	\end{subfigure}
	\caption{Computed total drag on a straight parallel filament moving along
	a boundary parallel to its tangent, here with $\epsilon/L=10^{-2}$.
	\subref{fig:results:drag_compare_uniform_filament:close} Near the boundary
	we see very good agreement between the regularised Stokeslet segment
	(\rss{}) solution (dot-dashed) and the wall-corrected resistive force
	theory (\wrft{}) solution of \citeauthor{Katz1975} (solid), with fair
	agreement maintained outside the region of validity of \wrft{} as
	$h/L\sim1$. The boundary element method (BEM) data of
	\citeauthor{Ramia1993} is marked with squares, adapted from \citet[Fig
	4c]{Ramia1993} and showing good agreement with the regularised Stokeslet
	segments.
	\subref{fig:results:drag_compare_uniform_filament:far} Similar agreement
	can be seen far from the boundary, with the asymptotic solution of
	\citeauthor{Brenner1962} losing validity when $h/L\sim1$ though remaining
	within 10\% of the RSS solution.
	\label{fig:results:drag_compare_uniform_filament}}
\end{figure}

% Agreement with corrected RFT, and thus with Ramia's BEM. Wall RFT is ok in the far field too (previously untested)
For the case of a straight uniform filament aligned parallel to a planar
boundary, utilising the approach of \cref{sec:methods:force_given_kinematics}
we compute the hydrodynamic drag on the slender body as it moves parallel to
the wall along its tangent at unit non-dimensional velocity, comparing the
solutions given by the methods of regularised Stokeslet segments, free-space
RFT and wall-corrected RFT. Having normalised by the
\frft{} solution, it being independent of the boundary separation $h$, we show
the computed non-dimensional total drag in
\cref{fig:results:drag_compare_uniform_filament}. Good agreement near the
boundary can be seen between the \wrft{} of \citeauthor{Katz1975} and our
implementation of the method of regularised Stokeslet segments, the former as
previously validated in the limit $h/L \rightarrow0$ with high-accuracy
boundary element methods by \citet{Ramia1993}. Indeed, direct comparison of
the \rss{} solution with \citet[fig. 4c]{Ramia1993}, where we note that here
we are taking $\epsilon/L=10^{-2}$ to ensure the filament radius matches that
used in the boundary element computations, which thus provide additional
verification of our methodology. Further, far from the boundary the correction
of \citet{Brenner1962} lies within 10\% of the RSS solution, evidencing good
overall agreement between the two schemes as the normalised wall separation
$h/L$ increases. Analogous agreement between these methodologies can
additionally be seen for total normal drag when moving normal to the boundary,
thus we conclude that the corrections to resistive force theory of
\citeauthor{Brenner1962} and \citeauthor{Katz1975} agree closely with
regularised Stokeslet segments for straight filaments in their respective
asymptotic regimes. Hence, agreement in more generality may be reasonably
hypothesised.

\subsection{Computing hydrodynamic drag on filaments parallel to boundaries}
\label{sec:results:parallel}
% Report that even wall-corrected RFT is unreliable for this.
Following the established good agreement of \wrft{} and \rss{} for straight
filaments we proceed to evaluate the agreement for curved filaments moving in
a plane parallel to the no-slip boundary. Motivated by the previous use of
free-space resistive force theory for drag determination from captured
kinematic data \citep{Ooi2014,Gaffney2011,Ishijima2011,Friedrich2010}, we
compute the forces on a free filament from the smoothed kinematic data of
\citet{Ishimoto2017b} corresponding to the flagellum of a human spermatozoon

\begin{figure}
\centering
	\begin{subfigure}[c]{0.45\textwidth}
		\centering
		\includegraphics[width=\textwidth]{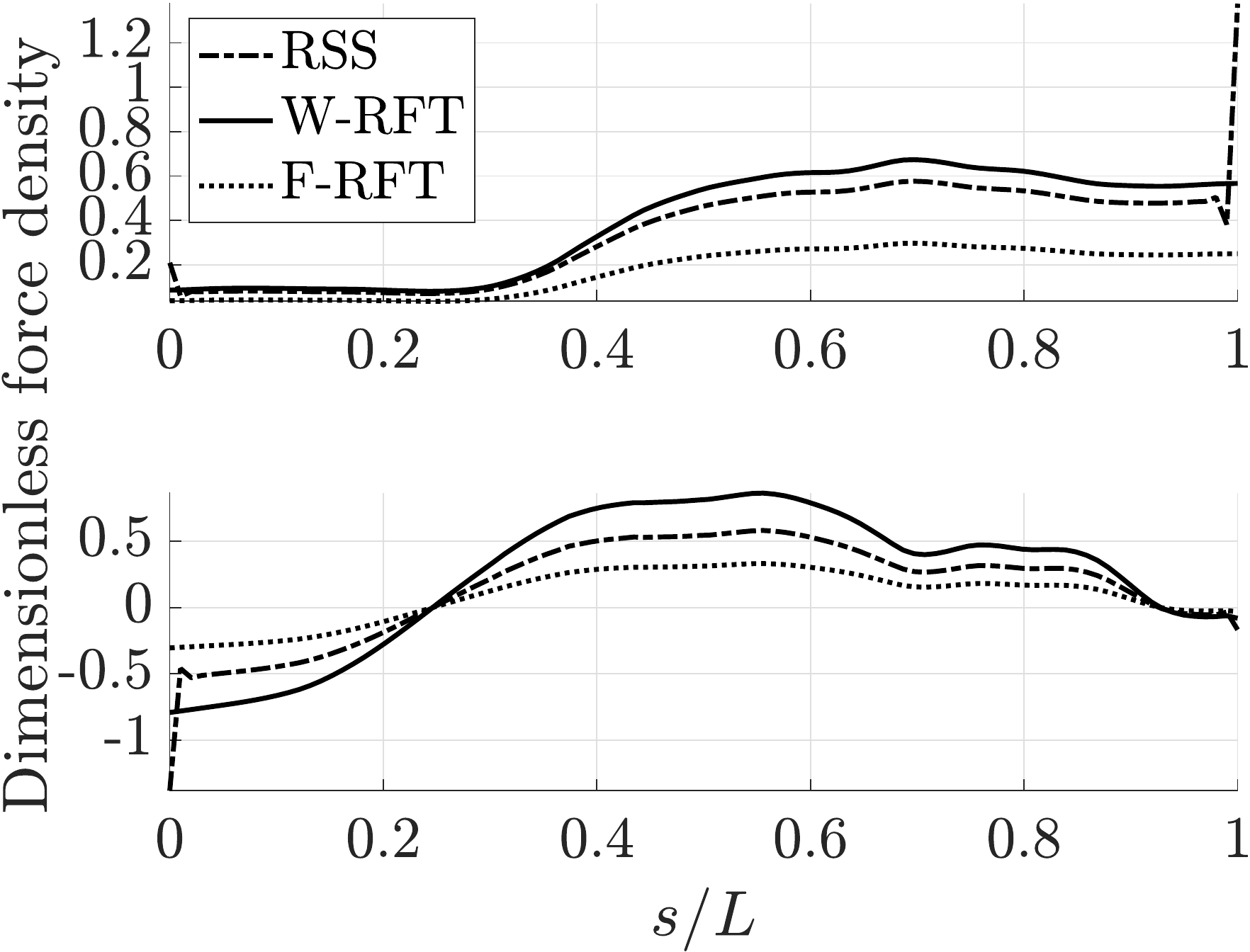}
		\caption{\label{fig:results:drag_compare_parallel_sperm:single_frame}}
	\end{subfigure}
	\begin{subfigure}[c]{0.45\textwidth}
		\centering
		\includegraphics[width=\textwidth]{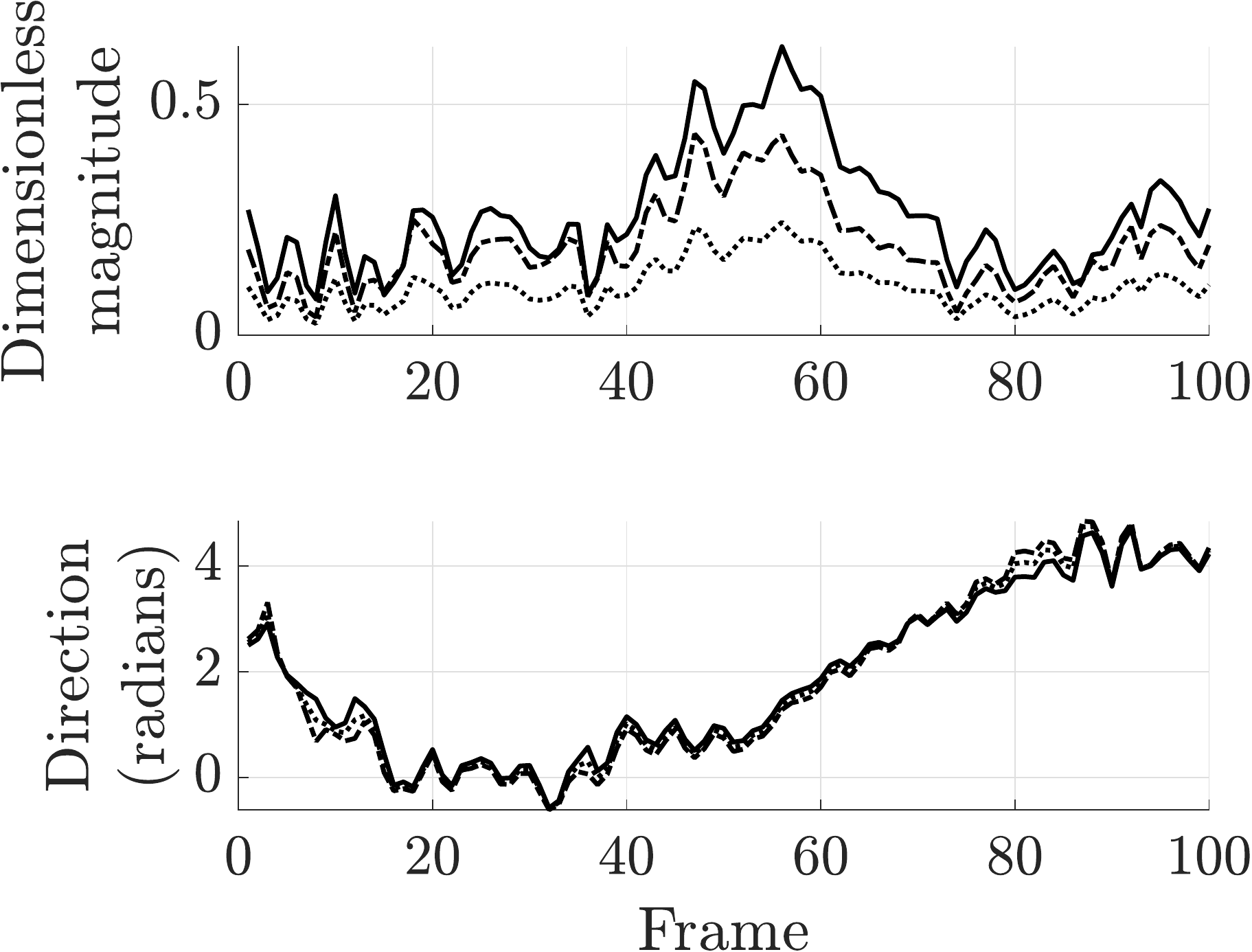}
		\caption{\label{fig:results:drag_compare_parallel_sperm:force_head}}
	\end{subfigure}

	\begin{subfigure}[c]{0.45\textwidth}
		\centering
		\includegraphics[width=\textwidth]{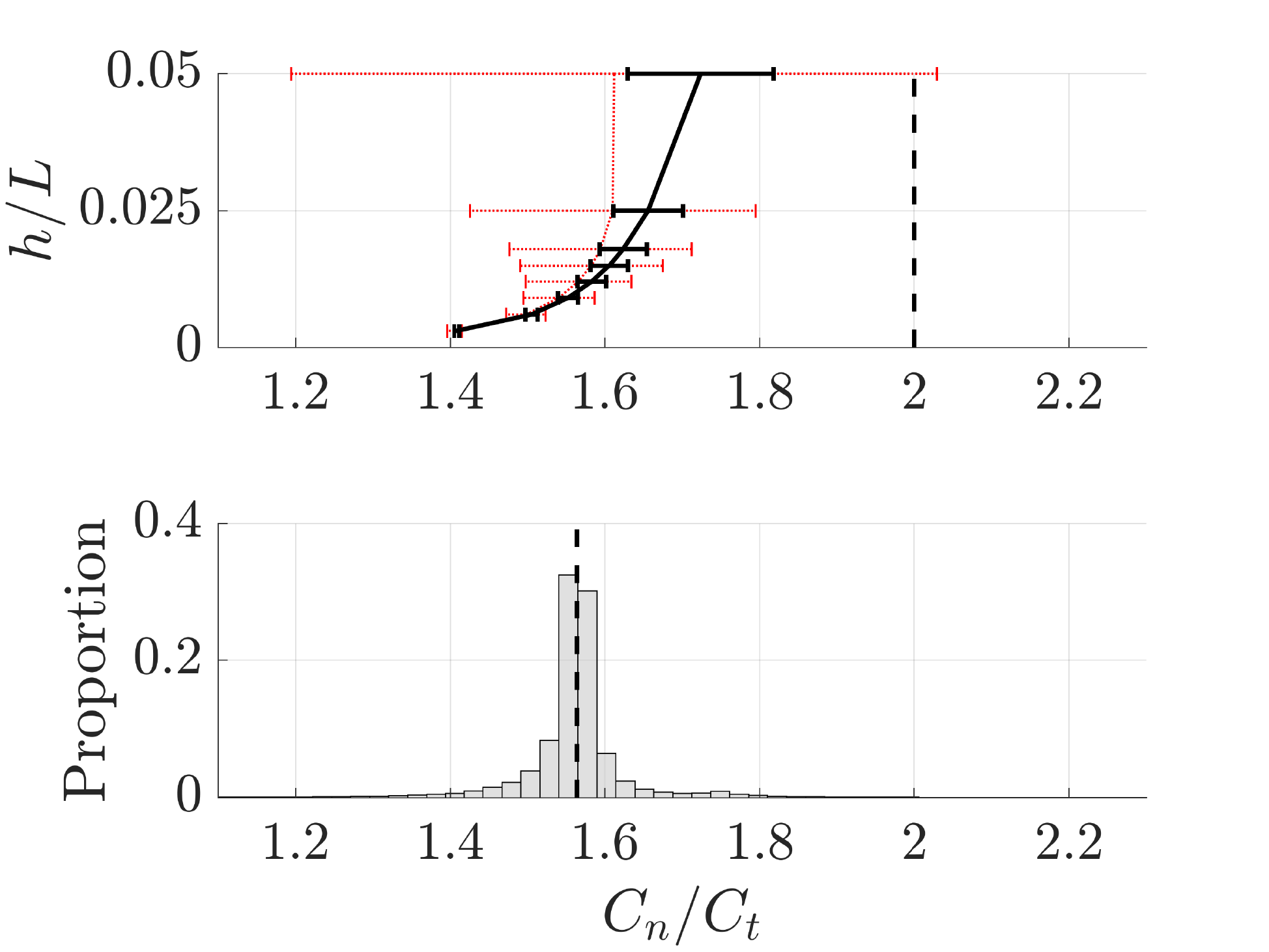}
		\caption{\label{fig:results:drag_compare_parallel_sperm:coeffs}}
	\end{subfigure}	
	\begin{subfigure}[c]{0.45\textwidth}
		\centering
		\includegraphics[width=\textwidth]{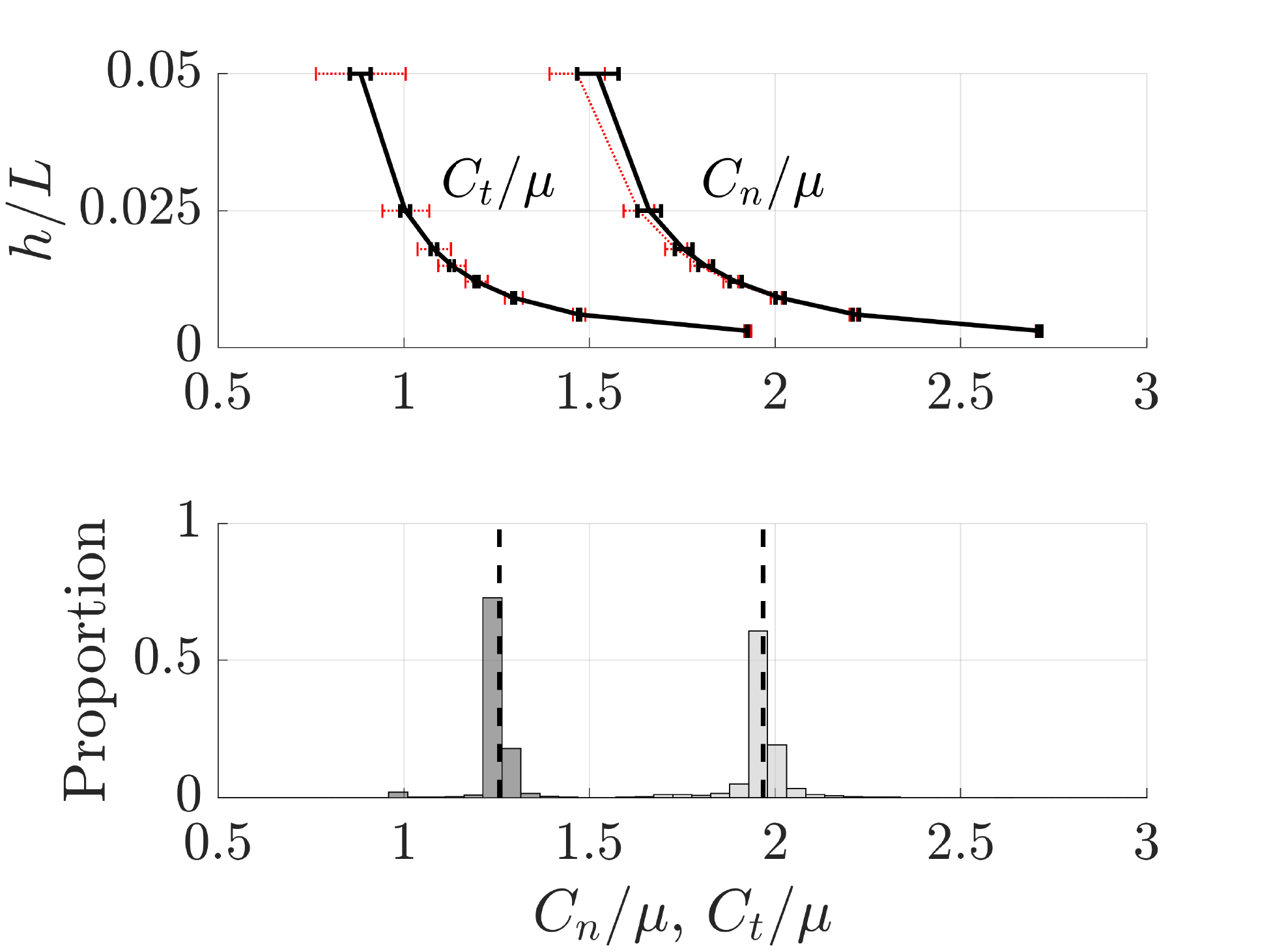}
		\caption{\label{fig:results:drag_compare_parallel_sperm:coeffs_individual}}
	\end{subfigure}
\caption{Drag calculations from kinematic data captured parallel to a planar
boundary, with data from \citet{Ishimoto2017b}.
\subref{fig:results:drag_compare_parallel_sperm:single_frame} The
tangential and normal components of the drag along the filament for a single
frame (upper and lower panels respectively) at separation $h/L=0.01$, showing
good qualitative agreement between methodologies of drag computation and fair
quantitative agreement between \wrft{} and \rss{}.
\subref{fig:results:drag_compare_parallel_sperm:force_head} The magnitude
and direction of the total integrated force on the filament (upper and lower
panels respectively) at separation $h/L=0.01$, with direction measured
relative to an arbitrary fixed axis. As in
\subref{fig:results:drag_compare_parallel_sperm:single_frame}, agreement
is fair and qualitative features are captured by all methodologies.
Remarkably, the direction of the resultant force as computed using the
resistive force theories strongly agrees with that of the non-local \rss{}.
Endpoint effects are visible in the upper panel of
\subref{fig:results:drag_compare_parallel_sperm:single_frame}, as
discussed in detail in \cref{sec:methods:endpoint_effects}.
\subref{fig:results:drag_compare_parallel_sperm:coeffs} Ratio of effective drag
coefficients $C_n/C_t$, as computed via \rss{}, against boundary separation
(upper), shown as the median over all frames and all material points (740
frames, 100 points per frame). The ratio as predicted by
\citeauthor{Katz1975}, 2, is shown as a dashed line for comparison. Errorbars
corresponding to half the interquartile range are shown, representative of
dispersal about the median as the distributions do not appear significantly
skewed. Analogous results for artifically-pinned kinematic data are shown in
red (dotted, thin), highlighting a reduction in validity of resistive force
theories for pinned data but retaining notable accuracy when very close to the
boundary. Lower panel is a histogram corresponding to $h/L=0.01$, showing the
distribution of effective coefficient ratios in the free-swimming case,
highlighting tight grouping about the median.
\subref{fig:results:drag_compare_parallel_sperm:coeffs_individual} Median
values of effective resistive coefficients against boundary separation
(upper), with errorbars of half the interquartile range. Artificially-pinned
data is shown in red (dotted, thin) for comparison. Lower panel is histogram
corresponding to $h/L=0.01$ (lower), showing the tight distribution of
dimensionless effective drag coefficients $C_n/\mu$ and $C_t/\mu$, with
$C_t/\mu$ shown darker. Medians are shown dashed.}
\label{fig:results:drag_compare_parallel_sperm}
\end{figure}

We first consider the filament moving in very close proximity to the boundary,
with normalised separation $h/L = 0.01$, within the typical range of
slithering motion for spermatozoa \citep{Nosrati2015}. A summary of the
resulting drag calculations is shown in
\cref{fig:results:drag_compare_parallel_sperm}, from which we observe that the
wall-corrected RFT of \citet{Katz1975} can demonstrate strong pointwise
agreement with regularised Stokeslet segments over a single frame
(\cref{fig:results:drag_compare_parallel_sperm:single_frame}). However,
calculations of the total drag force over the filament highlight a general
trend of over-estimation of force densities by \wrft{}, with the differences
between \wrft{} and \rss{} having a median of 37\% over the 100-frame range
shown here, with differences measured relative to the \rss{} value. Free-space
resistive force theory appears to do the opposite, systematically
underestimating the magnitude of total drag, with increased median deviation
of 45\% from the \rss{} solution. Most notable however is the tight clustering
of the effective ratio of drag coefficients $C_n/C_t$, shown in
\cref{fig:results:drag_compare_parallel_sperm:coeffs}, computed pointwise from
the \rss{} solution and with an analogous distribution of the effective values
of $C_n/\mu$ and $C_t/\mu$ shown in
\cref{fig:results:drag_compare_parallel_sperm:coeffs_individual}. This
suggests that an appropriate choice of resistive coefficient would enable the
accurate approximation of local filament drag using only a local theory, in
concurrence with the findings of \citet{Friedrich2010} though seen here much
closer to the boundary and at a local level, rather than
\citeauthor{Friedrich2010}'s comparison to observed sperm behaviour, which
necessarily averages over the cell. We verify this result using additional
waveform data, modifying the kinematic data of \citet{Ishimoto2017b} to
crudely approximate a pinned spermatozoa by fixing both the endpoint and local
tangent in space, along with the idealised pinned waveforms of
\citet{Curtis2012}. In particular, the tight clustering of effective
coefficient ratios at low boundary separations is retained, though we note
reduced agreement compared to that present for free-swimming data.
Surprisingly, these coefficients and ratios do not align with the corrected
coefficients of \citet{Katz1975}, derived for straight filaments parallel to a
boundary.

Additionally, for approximate separations $0.3<h/L<1$ we observe very strong
agreement between resistive force theories and the non-local solution, with
median differences in total drag between methodologies consistently less than
15\%, in many cases being less than 5\%. This is coupled with additional
agreement concerning the direction of the resultant filament drag, which is
retained even at reduced separations
(\cref{fig:results:drag_compare_parallel_sperm:force_head}, lower). However,
lost at such intermediate separations is any clear choice of effective
resistive coefficient ratio, with the distribution of effective ratios
significantly broadening above $h/L\approx0.05$.

At a much greater distance from the boundary, with $h/L=10$, as expected
\frft{} and \wrft{} give approximately equal estimates for the drag on the
free-swimming filament, with the \wrft{} solution having approached that of
\frft{}. As in the case of near-boundary swimming, only small differences are
present between the \wrft{} and \rss{} solutions, with median differences
between methods of around 6\%, in this case with the magnitude of all computed
drag forces having been reduced from their near-wall values by approximately a
factor of two. Thus, in the medium and far-field of a boundary resistive force
theories appear remarkably accurate for determining the total drag on even
curved filaments moving parallel to the boundary, though surprisingly at this
increased boundary separation there is little grouping of the effective
resistive coefficient ratios.

% \begin{figure}
% \centering
% 	\begin{subfigure}[c]{0.45\textwidth}
% 		\centering
% 		\includegraphics[width=\textwidth]{figs/drag_compare_parallel_beat_plane/pointwise_drag_single_frame_medium-eps-converted-to.pdf}
% 		\caption{\label{fig:results:drag_compare_parallel_sperm_far:single_frame}}
% 	\end{subfigure}
% 	\begin{subfigure}[c]{0.45\textwidth}
% 		\centering
% 		\includegraphics[width=\textwidth]{figs/drag_compare_parallel_beat_plane/total_force_medium-eps-converted-to.pdf}
% 		\caption{\label{fig:results:drag_compare_parallel_sperm_far:force_head}}
% 	\end{subfigure}
% \caption{Drag calculations from kinematic data captured parallel to a planar
% boundary for separation $h/L=10$.
% \subref{fig:results:drag_compare_parallel_sperm:single_frame} The
% tangential and normal components of the drag along the filament for a single
% frame (top and bottom panels respectively), showing good qualitative agreement
% between methodologies of drag computation but poor quantitative agreement.
% \subref{fig:results:drag_compare_parallel_sperm:force_head} The magnitude
% and direction of the total integrated force on the filament (top and bottom
% panels respectively), with direction measured relative to some fixed axis. As
% in \subref{fig:results:drag_compare_parallel_sperm:single_frame},
% agreement is poor but qualitative features are captured by all methodologies.
% Remarkably, the direction of the resultant force as computed using the
% resistive force theories strongly agrees with that of the non-local \rss{}.\label{fig:results:drag_compare_parallel_sperm_far}}
% \end{figure} 

\subsection{Hyperactivation-induced tugging of tethered spermatozoa}
\label{sec:results:pinned}
% We confirm the hypothesis of tugging from hyperactivated sperm when tethered, made initially using RFT, though we note that gross cancellation occurs and overall agreement is likely serendipitous.
Explored initially by \citet{Curtis2012} using both free-space and
wall-corrected resistive force theory, and reconsidered by \citet{Simons2014}
using regularised Stokeslets, we re-evaluate the observation that the
hyperactivation of mammalian spermatozoa aids in surface escape via a
beat-induced tugging effect on the tether point of boundary-attached
spermatozoa, and consider in detail the drag on the filament. Adopting the
idealised beat patterns used by \citeauthor{Curtis2012}, appropriately
non-dimensionalised, we position the base of the filament $0.01L$ from the
boundary and compute both the total and local drag from this kinematic data
for both hyperactivated and normal beating, noting that the plane of filament
beating is perpendicular to the boundary. Mirroring the setup of
\citeauthor{Curtis2012}, we assume that the filament is clamped at its base,
implemented by rotating the kinematic data to align the basal tangent in each
instant at some angle $\theta_0$ to the boundary.

\begin{figure}
\centering
	\begin{subfigure}[c]{0.32\textwidth}
		\centering
		\includegraphics[width=\textwidth]{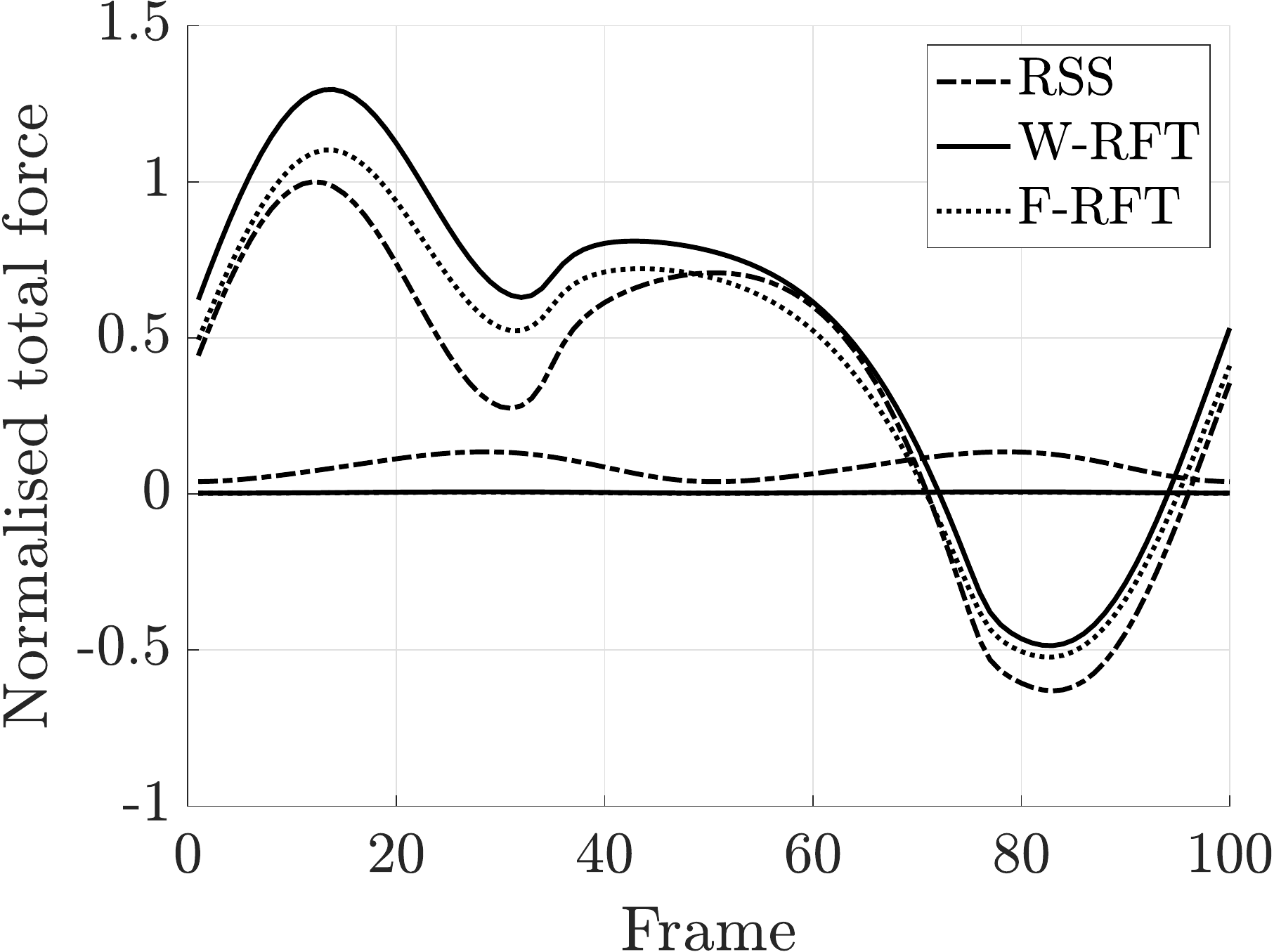}
		\caption{\label{fig:results:pinned_sperm:total_tugging}}
	\end{subfigure}
	\begin{subfigure}[c]{0.32\textwidth}
		\centering
		\includegraphics[width=\textwidth]{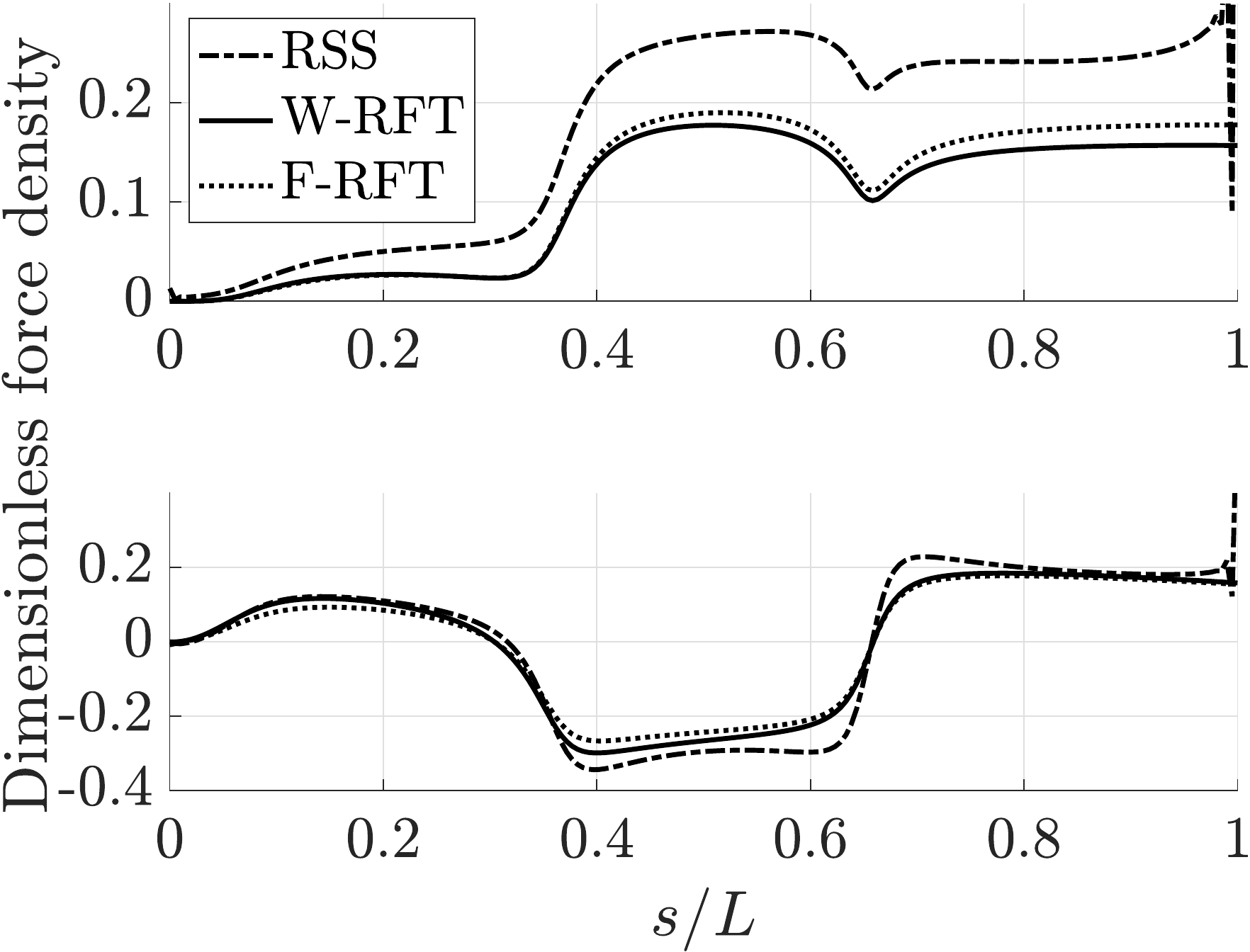}
		\caption{\label{fig:results:pinned_sperm:pointwise}}
	\end{subfigure}
	\begin{subfigure}[c]{0.32\textwidth}
		\centering
		\includegraphics[width=\textwidth]{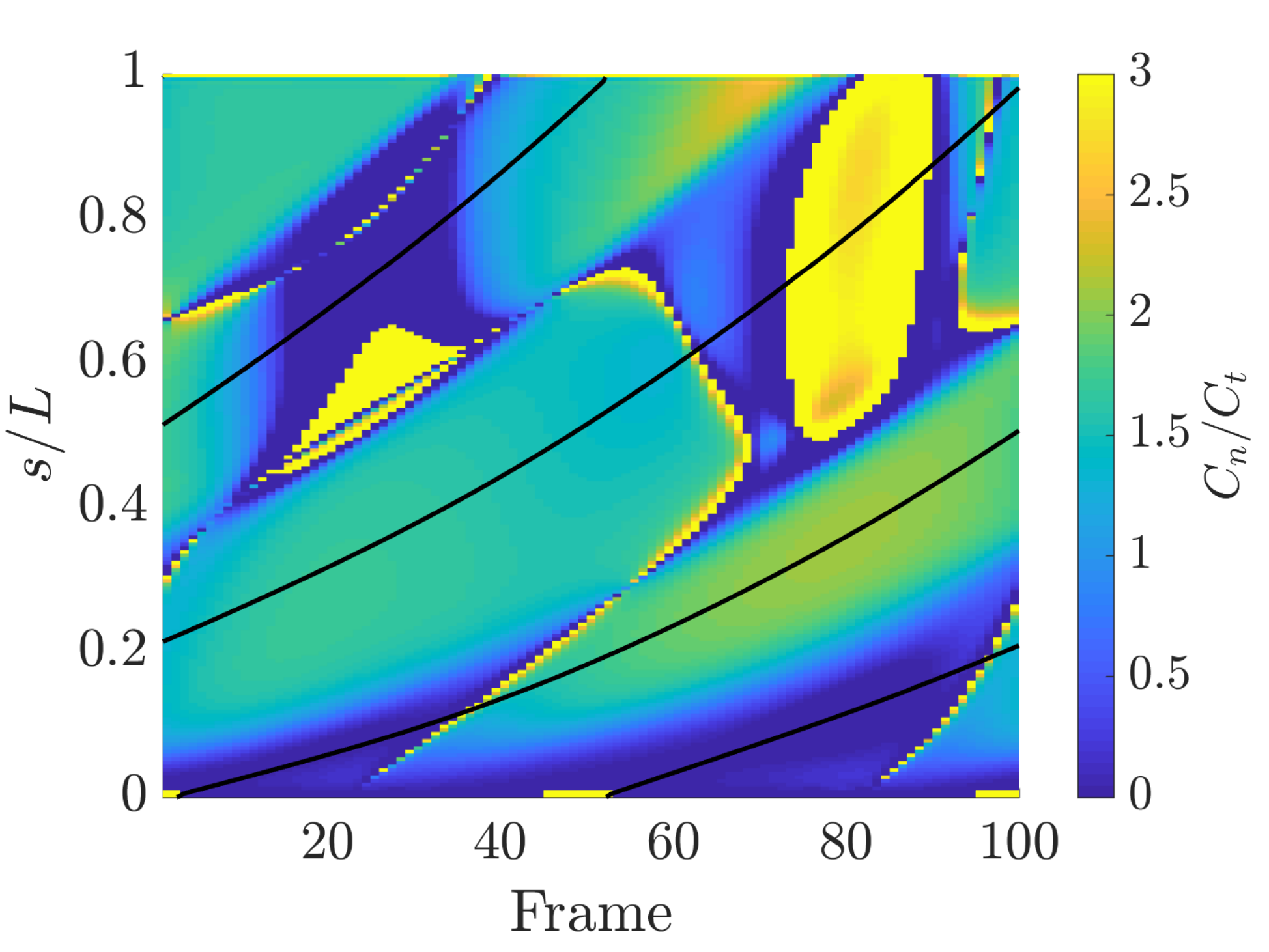}
		\caption{\label{fig:results:pinned_sperm:coeffs}}
	\end{subfigure}
	\caption{Drag computation for pinned spermatozoa with a base clamped
	perpendicular to a boundary over a single period of 100 frames.
	\subref{fig:results:pinned_sperm:total_tugging} Normalised total force on
	the boundary over a single beat period as computed by regularised
	Stokeslet segments (dot-dashed), free-space resistive force theory
	(dotted), and wall-corrected resistive force theory (solid), in the cases
	of hyperactive and normal beating patterns (high and low amplitudes
	respectively). Normal beating gives rise to a force on the boundary of
	unchanging sign, whilst hyperactivated beating generates a force whose
	sign changes over the beat period, corresponding to a so-called tugging
	effect. \subref{fig:results:pinned_sperm:pointwise} Tangential (upper
	panel) and normal components (lower panel) of force density as a function
	of arclength for a single frame, shown for the hyperactivated beat
	pattern. In contrast to the fair agreement between methods seen in
	\subref{fig:results:pinned_sperm:total_tugging}, with mean error of 24\%
	for the hyperactivated beat between \frft{} and \rss{}, tangential force
	density differs on average by 43\% between methods for the frame shown.
	\subref{fig:results:pinned_sperm:coeffs} Ratio between effective drag
	coefficients computed using regularised Stokeslet segments, with contours
	of zero filament curvature superimposed as black curves, showing
	significant variation in the effective ratio. Colour online.
	\label{fig:results:pinned_sperm}}
\end{figure}

\Cref{fig:results:pinned_sperm} shows the results of the drag computation over
a single beat period for both the hyperactivated and normal beat patterns for
$\theta_0=\pi/2$. We see reaffirmed by the method of regularised Stokeslet
segments the conclusion of \citeauthor{Curtis2012}, with the hyperactivated
beat pattern exhibiting a change of sign in force component perpendicular to
the boundary, whilst no such change is observed for the normally-beating
flagellum. Sampling $\theta_0\in[\pi/4,\pi/2]$, we note that this observation
holds across a range of basal orientations for these beat patterns, in
agreement with the RFT-established conclusions of \citeauthor{Curtis2012}.
Overall, \cref{fig:results:pinned_sperm:total_tugging} demonstrates fair
agreement between the local free-space resistive force theory solution and
that obtained using regularised Stokeslet segments, suggesting a surprising
validity in using simple local theories in this circumstance, as concluded
broadly by \citeauthor{Simons2014}. However, the pointwise values of
the drag density highlight a stark disagreement between local and non-local
theories, with deviations of up to 43\% for the tangential component of the
drag for the frame of hyperactivated beating shown in
\cref{fig:results:pinned_sperm:pointwise}. For reference, the integrated total
drag along the filament on average gives differences of only 24\% between
\rss{} and \frft{}, suggesting that a serendipitous cancellation occurs when
integrating over the flagellum. 

In \cref{fig:results:pinned_sperm:coeffs} we show the ratio of effective
tangential and normal drag coefficients for the hyperactivated beating, along
with contours of zero curvature for reference. Significant variation in this
ratio can be seen, in stark contrast to those computed for near-wall parallel
swimming above. Thus the conclusion of \citet{Friedrich2010}, that resistive
force theory is accurate to high precision, does not hold for a bound
spermatozoon, and hence the use of simple resistive force theories for
near-boundary drag calculations is not reliable in general. This precludes the
general use of the near-boundary correction of \citet{Katz1975} in this
circumstance, supported by the poor agreement seen in
\cref{fig:results:pinned_sperm}, whilst the far-field result of
\citet{Brenner1962} is inappropriate in such close boundary proximity.

\subsection{Morphological bifurcation of cantilevered filaments in an
oscillatory flow}\label{sec:results:oscillatory_morphological} In the absence
of a resistive force theory capable of accurately quantifying the details of
pinned filament motion perpendicular to boundaries, we extensively utilise the
non-local theory of regularised Stokeslet segments to consider the
fully-coupled elastohydrodynamics of a cantilevered filament in an oscillating
background flow, as formulated in \cref{sec:methods:elastohydrodynamic}. With
a planar infinite no-slip boundary situated at $y=0$, we consider the
non-dimensional background flow with velocity $\hat{\vec{u}}_b =
\hat{a}\sin{\left(2\pi
\hat{t} - \hat{t}_0\right)}\hat{y}\ex$ for amplitude $\hat{a}$ and phase
$\hat{t}_0$, giving rise to the modified non-dimensional system
\begin{equation}\label{eq:results:oscillatory_morphological:full_system}
	-\Eh{} \hat{B}\hat{A}^{-1}\hat{Q}\dot{\hat{\vec{\theta}}} = \hat{\vec{R}} - \Eh{} \hat{B}\hat{A}^{-1}\hat{\vec{U}}_b\,,
\end{equation}
where the components of $\hat{\vec{U}}_b$ are given by the background flow
evaluated at the endpoints of filament segments, explicitly
\begin{equation}
	\hat{\vec{U}}_b = \hat{a}\sin{\left(2\pi \hat{t}-\hat{t}_0\right)}\left(\hat{y}_1,0,\hat{y}_2,\ldots,\hat{y}_{N+1},0\right)^T\,.
\end{equation}
Noting that $\vec{n}(s_1)$ and $\vec{m}(s_1)$ are a priori unknown for a
cantilevered filament, we impose the appropriate constraints of
$\dot{x}_1=\dot{y_1}=\dot{\theta}_1=0$ in place of the total force and
moment-balance equations, again yielding a square linear system. Under the
assumption of a straight initial configuration with $\theta_i=\pi/2$ for
$i=1,\ldots,N$, which we make throughout, there is freedom in the three
non-dimensional parameters $\hat{a}$, $\hat{t}_0$ and $\Eh{}$, the latter
being equivalent to the sperm number $\Sp{}$. In \cref{app:phase} we consider
in detail the effects on the dynamics of varying the phase $\hat{t}_0$ of the
background flow, concluding that phase can effectively be neglected in
long-time dynamics, and thus we will fix $\hat{t}_0=0$, noting that this
breaks the inherent left-right symmetry of the initial condition.

\begin{figure}
	\centering
	\begin{subfigure}[c]{0.32\textwidth}
		\centering
		\includegraphics[width=\textwidth]{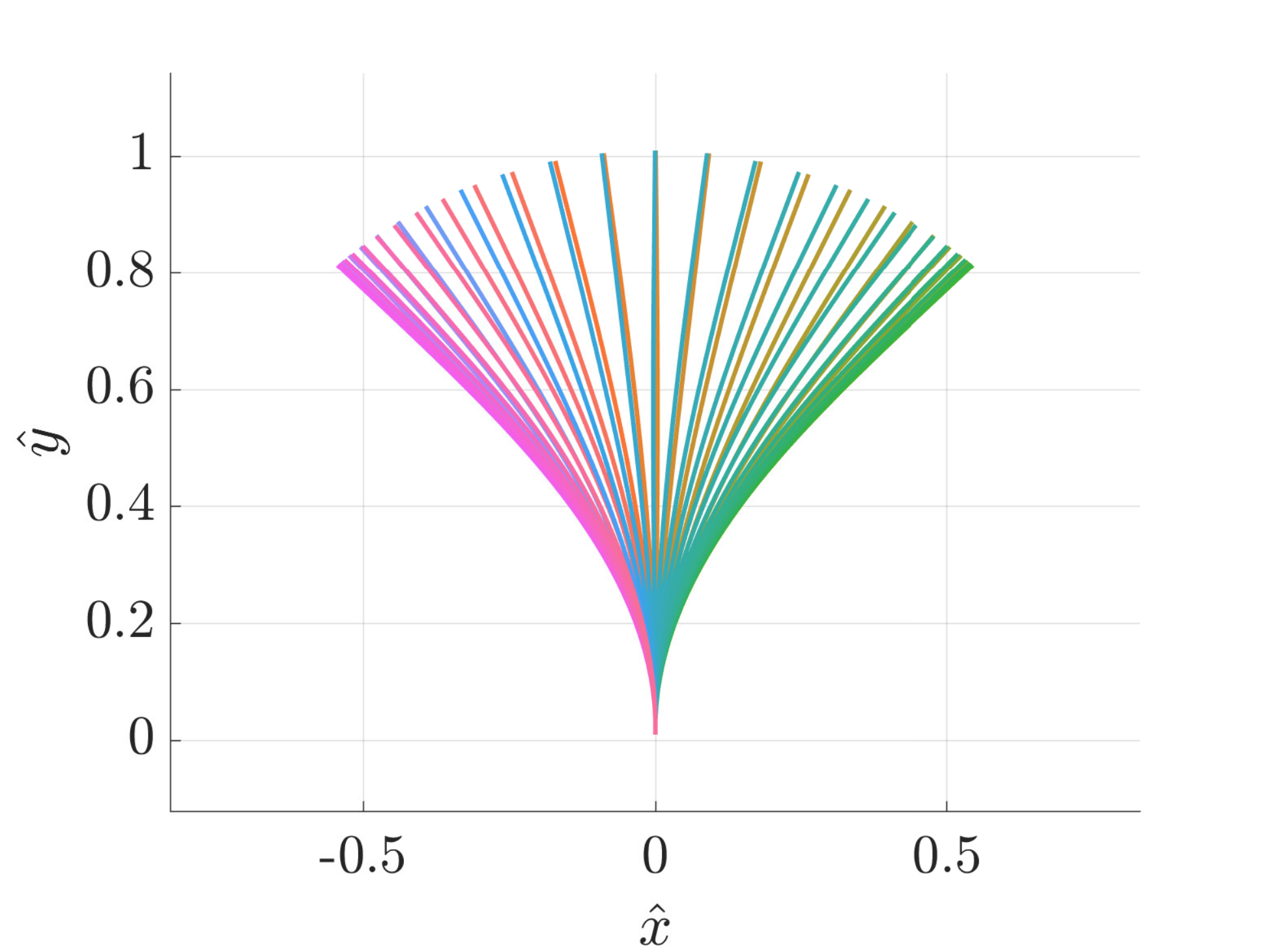}
		\caption{\label{fig:results:oscillatory_morphological:modes:Sp_low}}
	\end{subfigure}
	\begin{subfigure}[c]{0.32\textwidth}
		\centering
		\includegraphics[width=\textwidth]{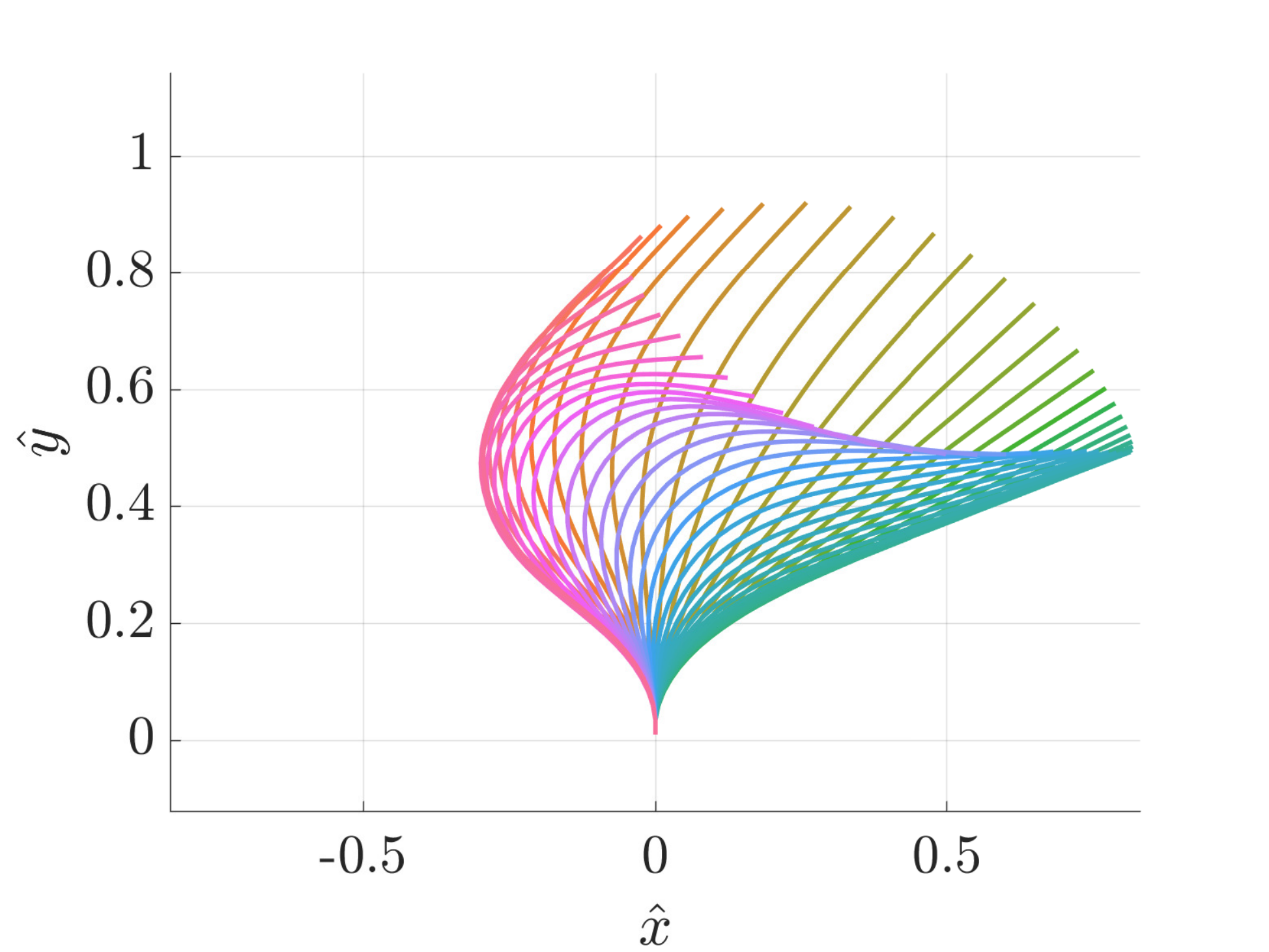}
		\caption{\label{fig:results:oscillatory_morphological:modes:Sp_med}}
	\end{subfigure}
	\begin{subfigure}[c]{0.32\textwidth}
		\centering
		\includegraphics[width=\textwidth]{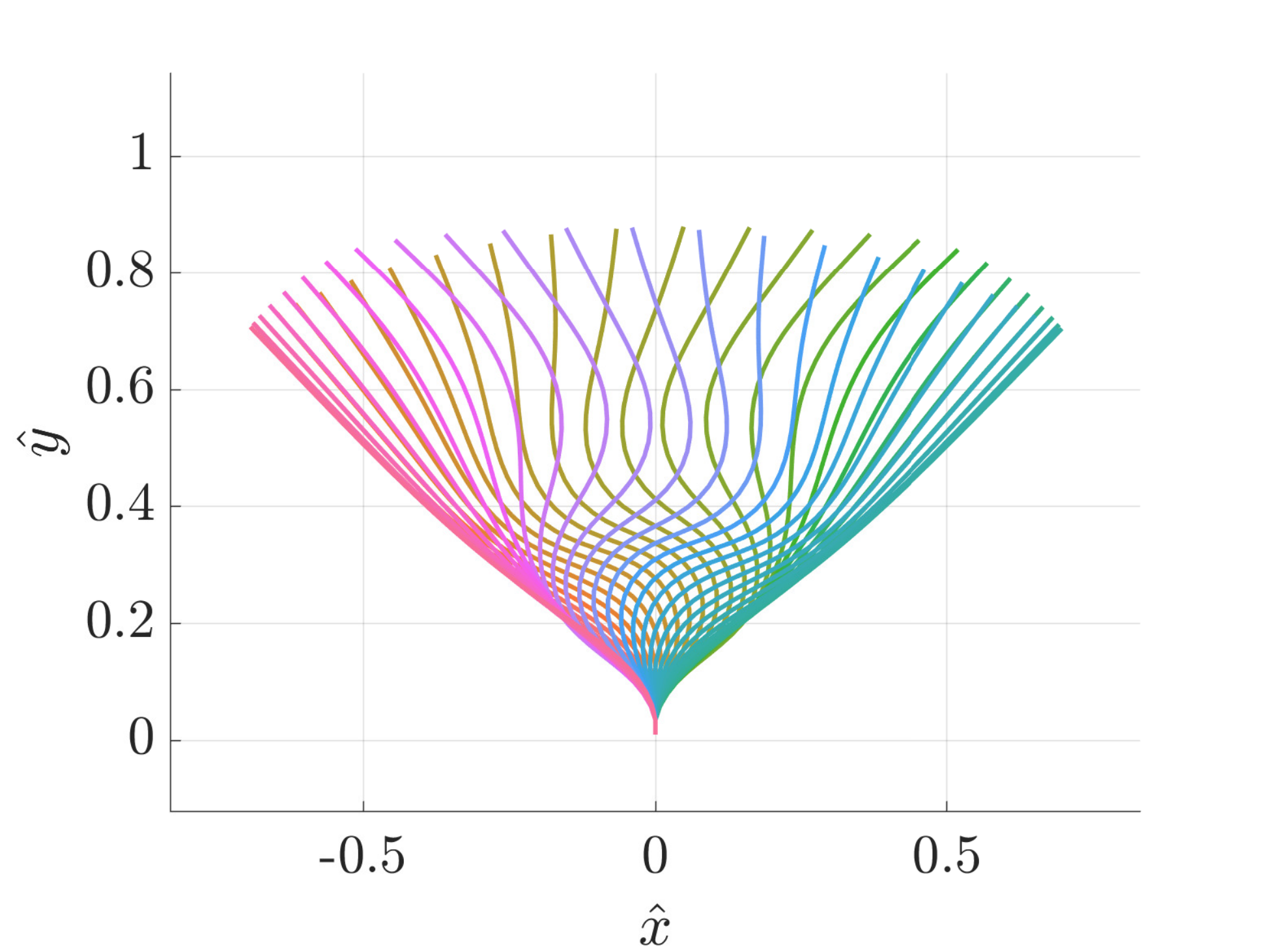}
		\caption{\label{fig:results:oscillatory_morphological:modes:Sp_high}}
	\end{subfigure}
	\begin{subfigure}[c]{0.32\textwidth}
		\centering
		\includegraphics[width=\textwidth]{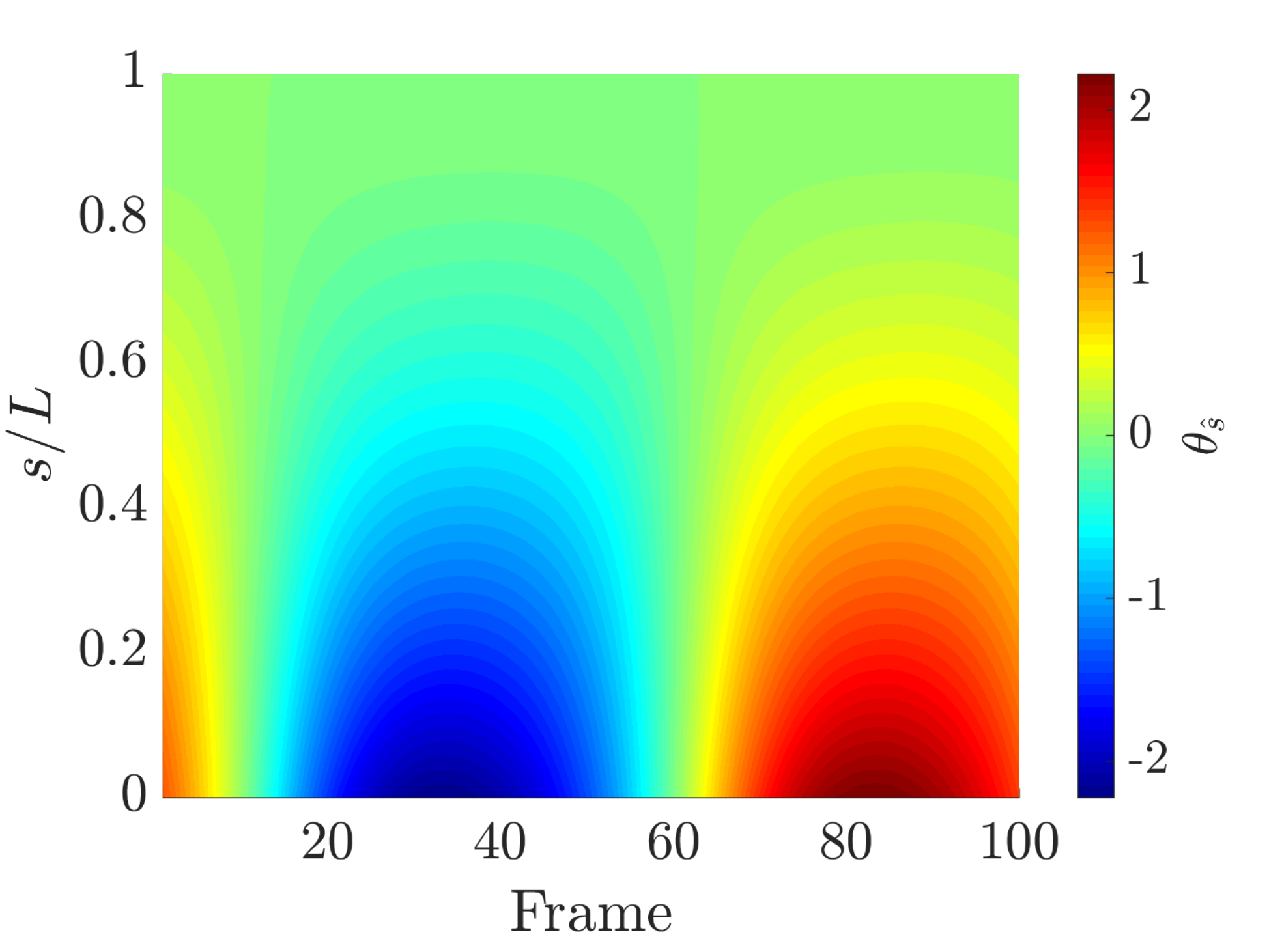}
		\caption{\label{fig:results:oscillatory_morphological:modes:Sp_low_curv}}
	\end{subfigure}
	\begin{subfigure}[c]{0.32\textwidth}
		\centering
		\includegraphics[width=\textwidth]{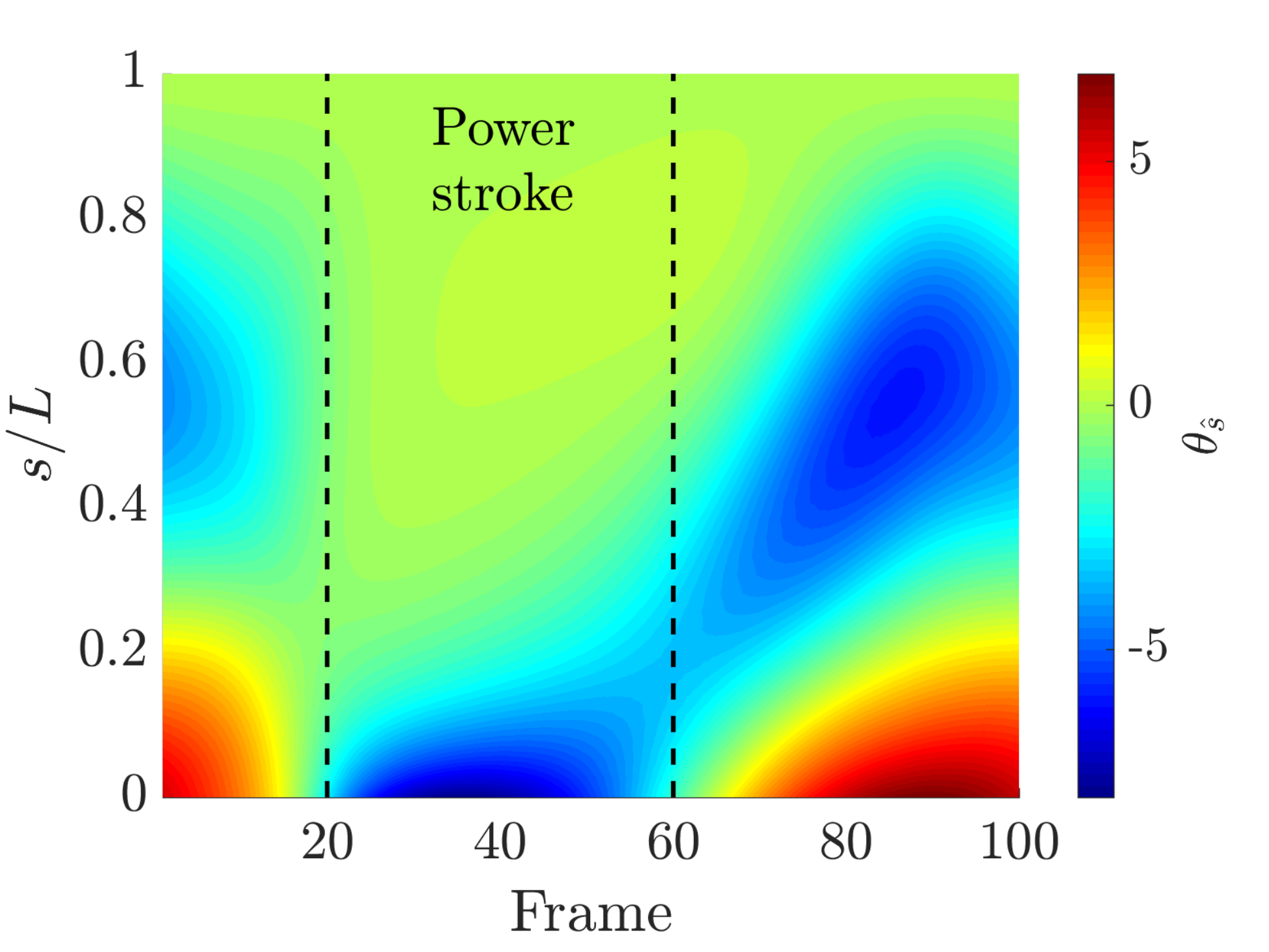}
		\caption{\label{fig:results:oscillatory_morphological:modes:Sp_med_curv}}
	\end{subfigure}
	\begin{subfigure}[c]{0.32\textwidth}
		\centering
		\includegraphics[width=\textwidth]{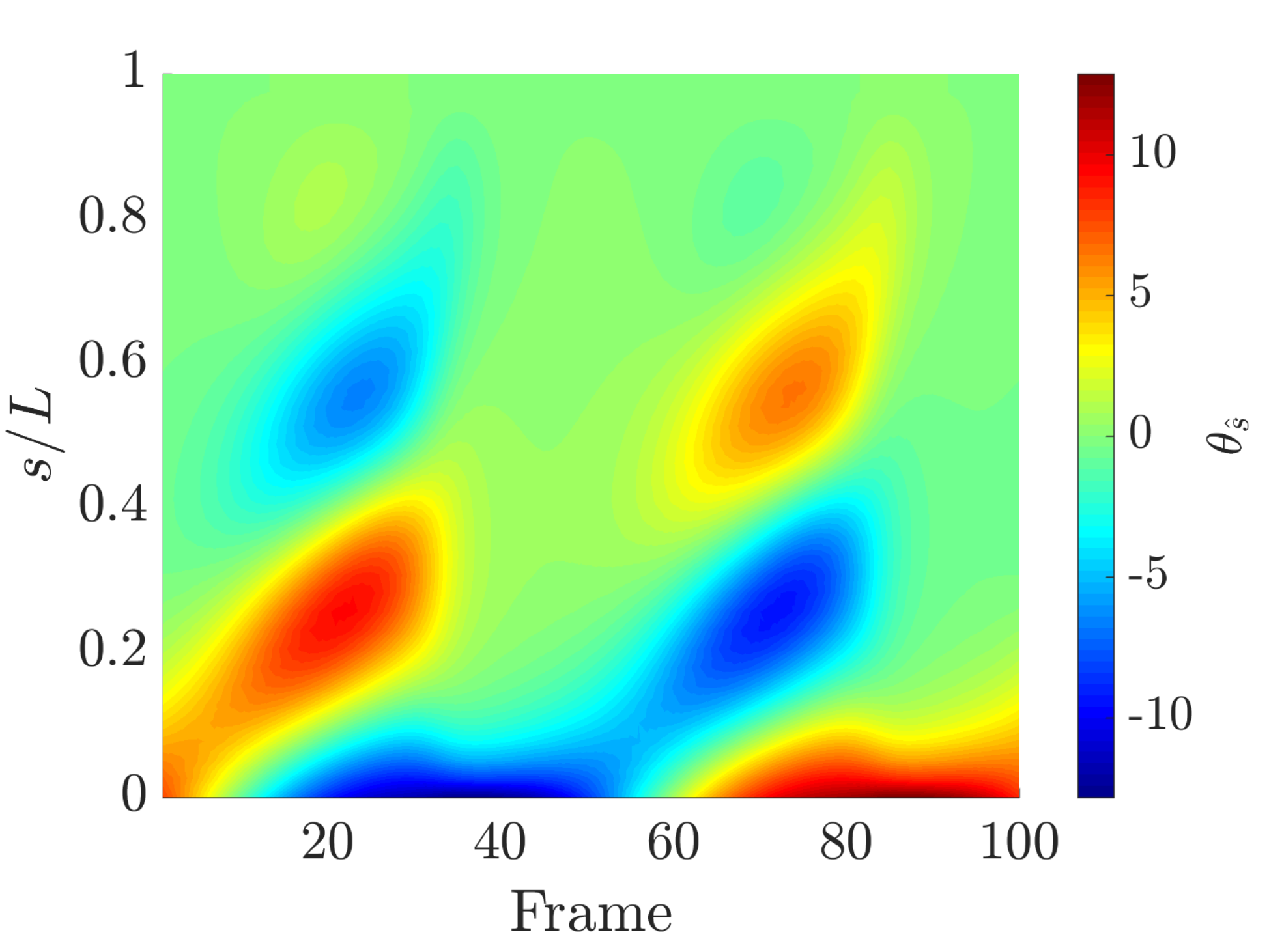}
		\caption{\label{fig:results:oscillatory_morphological:modes:Sp_high_curv}}
	\end{subfigure}
	\caption{Periodic beating of a cantilevered filament in oscillatory background
	flow of amplitude $\hat{a}=2\pi$.
	\subref{fig:results:oscillatory_morphological:modes:Sp_low} At low sperm
	numbers, here with $\Sp{}=1.14$, flow-driven beating is approximately
	symmetric.
	\subref{fig:results:oscillatory_morphological:modes:Sp_med} For medium
	sperm numbers a remarkable ciliary-type asymmetric beating emerges, shown
	here for $\Sp{}=2.84$ and globally stable, with a clear distinction
	between effective forward and reverse strokes.
	\subref{fig:results:oscillatory_morphological:modes:Sp_high} At higher
	sperm numbers the filament buckles significantly and regains a symmetric
	periodic beat, here with $\Sp{}=4.55$.
	\subref{fig:results:oscillatory_morphological:modes:Sp_low_curv}--\subref{fig:results:oscillatory_morphological:modes:Sp_high_curv}
	Plots of signed curvature corresponding to
	\subref{fig:results:oscillatory_morphological:modes:Sp_low}--\subref{fig:results:oscillatory_morphological:modes:Sp_high}
	above, sampled at 100 frames per period and each independently scaled for
	clarity. Symmetry is clearly visible in
	\subref{fig:results:oscillatory_morphological:modes:Sp_low_curv} and
	\subref{fig:results:oscillatory_morphological:modes:Sp_high_curv}, with
	the latter displaying a low-order buckling mode.
	\subref{fig:results:oscillatory_morphological:modes:Sp_med_curv}
	exemplifies cilia-like beating, with the effective power stroke occurring
	approximately between frames 20 and 60 (shown black, dashed). Colour
	online.
	\label{fig:results:oscillatory_morphological:modes}}
\end{figure}

For fixed $\hat{a}$ we consider the beating of the filament as it is driven by
the background flow with the sperm number being varied. With $\hat{a}=2\pi$,
we showcase in \cref{fig:results:oscillatory_morphological:modes} the diverse
range of observed behaviours. For low sperm numbers, in this case below a
threshold value of around $\Sp{}=2.5$, we observe symmetric, low amplitude
beating of the filament. Increasing the sperm number beyond this point results
in the remarkable emergence of an asymmetric beating, shown in
\cref{fig:results:oscillatory_morphological:modes:Sp_med}, strongly resembling
the characteristic beating of a cilium. In particular, we see reproduced in
this passive, flow-driven filament the defining power and recovery strokes of
beating cilia \citep{Brennen1977}.

\begin{figure}
	\centering
	\begin{subfigure}[c]{0.32\textwidth}
		\centering
		\includegraphics[width=\textwidth]{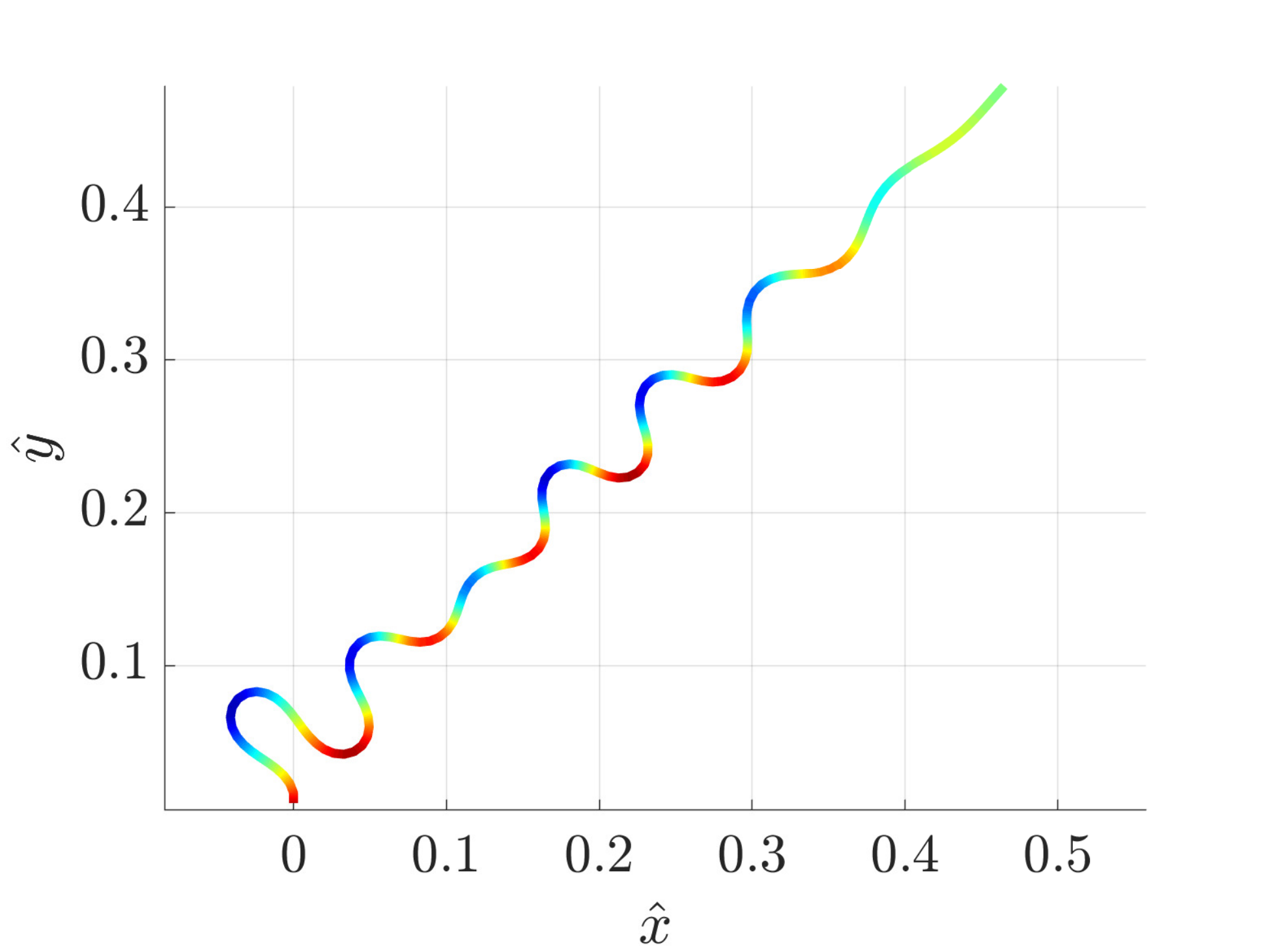}
		\caption{\label{fig:results:oscillatory_morphological:buckling:frame}}
	\end{subfigure}
	\begin{subfigure}[c]{0.32\textwidth}
		\centering
		\includegraphics[width=\textwidth]{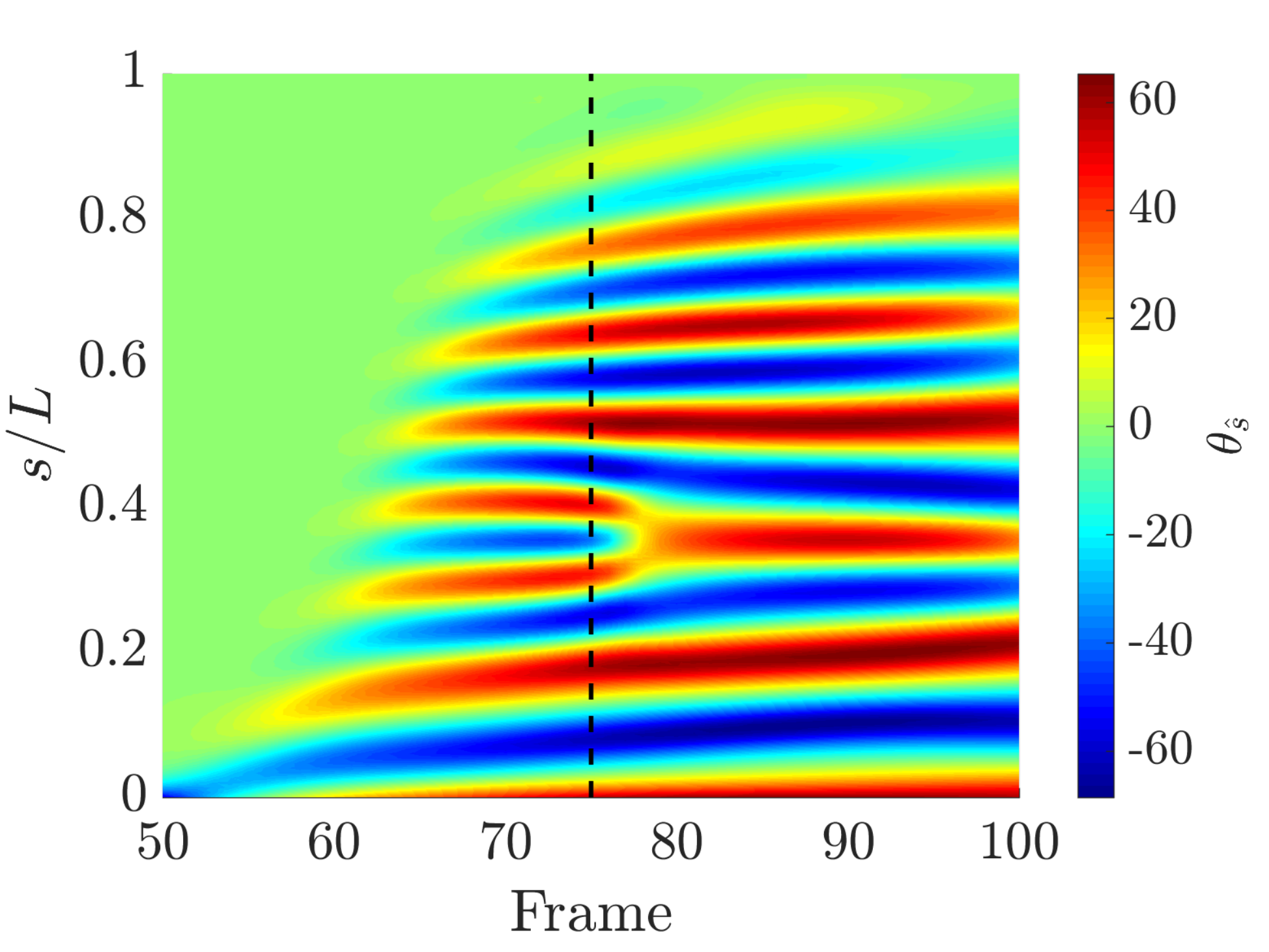}
		\caption{\label{fig:results:oscillatory_morphological:buckling:curv}}
	\end{subfigure}
	\begin{subfigure}[c]{0.32\textwidth}
		\centering
		\includegraphics[width=\textwidth]{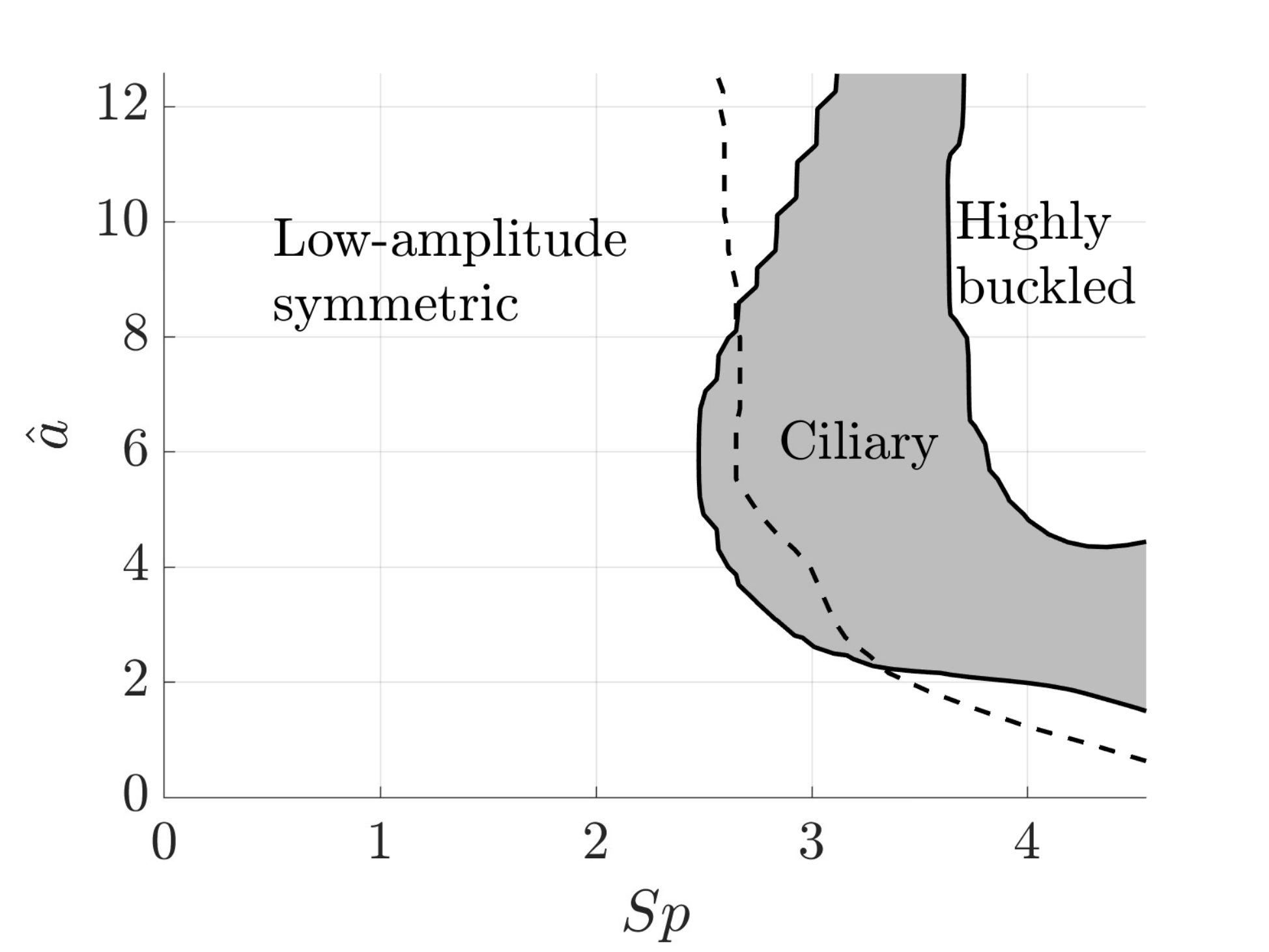}
		\caption{\label{fig:results:oscillatory_morphological:bifurcation}}
	\end{subfigure}
	\caption{Bifurcation into high order buckling modes for a filament with high sperm number,
	here with $\Sp{}=15$ and sampled at 100 frames over the initial period of
	the background flow.
	\subref{fig:results:oscillatory_morphological:buckling:frame} A single
	frame of the initial motion of the filament, where high-curvature regions
	can be seen to be resolved by the $N=150$ segments used here, with
	curvature represented by colour.
	\subref{fig:results:oscillatory_morphological:buckling:curv} The curvature
	of the filament over the latter part of the initial period of the
	background flow, from which we may identify the collapse of high order
	modes onto lower order buckling modes. The frame shown in
	\subref{fig:results:oscillatory_morphological:buckling:frame} is indicated
	on the curvature plot by a dashed line, with colour scalings consistent
	between plots. \subref{fig:results:oscillatory_morphological:bifurcation}
	A bifurcation diagram highlighting the regions of parameter space
	corresponding to low-amplitude symmetric, ciliary, and buckled modes of
	beating. Buckled modes appear to the right of the dividing dashed line, as
	defined by the transition between less than three and three regions of
	constant curvature sign along the filament. Furthermore, ciliary beating
	occurs within contours delimiting where $S$ is essentially unity, shown
	shaded. In the overlap region between ciliary and buckled beating we
	observe ciliary-type beats with buckled tips. Colour online.
	\label{fig:results:oscillatory_morphological:buckling}}
\end{figure}

Increasing the sperm number further results in the filament buckling and its
beating regaining approximate symmetry, as can be seen in
\cref{fig:results:oscillatory_morphological:modes:Sp_high} for $\Sp{}=4.55$.
For very high sperm numbers we see the appearance of buckling modes of higher
order, resolved in \cref{fig:results:oscillatory_morphological:buckling} using
$N=150$ segments to capture filaments with such exceptionally high curvatures.
With such high sperm numbers, we observe the collapse of modes onto those
corresponding to lower-order buckling as time progresses, in addition to a
greatly increased relaxation time to periodic beating in comparison to stiffer
filaments.

In order to classify the mode of periodic beating we introduce the symmetry
measure $S$, defined for our piecewise-linear filament with segment endpoint
coordinates $\hat{\vec{x}}_i(\hat{t})$ as 
\begin{equation}
	S = 1 - \frac{1}{2}\int\frac{\sum\limits_i \abs{\hat{\vec{x}}_i(\hat{t})+\hat{\vec{x}}_i(\hat{t}+1/2)}^2}{\sum\limits_i \abs{\hat{\vec{x}}_i(\hat{t})}^2}\intd{\hat{t}}\,,
\end{equation}
where the integral over $\hat{t}$ runs over half the period of the motion
after convergence to periodic beating has occurred. We note that the left-right
symmetric beats of
\cref{fig:results:oscillatory_morphological:modes:Sp_low,fig:results:oscillatory_morphological:modes:Sp_high}
give $S=1$, with motion that breaks left-right symmetry such as the distinct
forward and reverse strokes of ciliary beating yielding reduced values of $S$.
We find that values of $S$ a small tolerance less than unity indeed correspond
well to the ciliary-type beating of
\cref{fig:results:oscillatory_morphological:modes:Sp_med}. We additionally
distinguish between approximately-straight and highly buckled beating by
considering the number of filament regions with curvature of unchanging sign,
classifying filaments with three or more such regions as buckled. Shown in
\cref{fig:results:oscillatory_morphological:bifurcation} are the regions of
the \Spa{} parameter space that approximately correspond to low-amplitude
symmetric, ciliary, and buckled modes of filament beating.

\subsection{Time-averaged total drag on cantilevered filaments in oscillatory
flow corresponds to distinct beating
morphologies}\label{sec:results:oscillatory_drag} 
\begin{figure}
	\centering
	\begin{subfigure}[c]{0.45\textwidth}
		\centering
		\includegraphics[width=\textwidth]{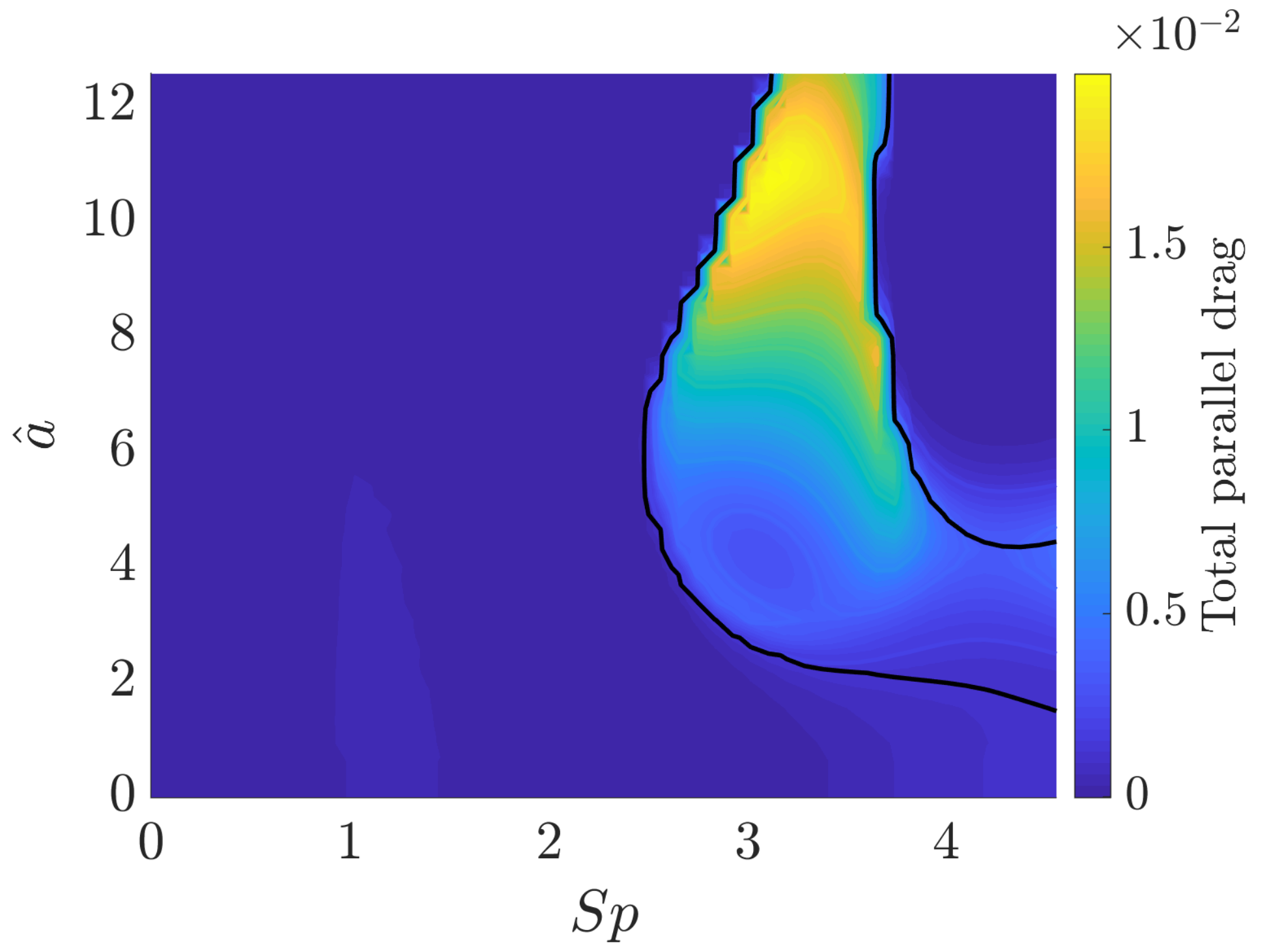}
		\caption{\label{fig:results:oscillatory_drag:drag_x}}
	\end{subfigure}
	\begin{subfigure}[c]{0.45\textwidth}
		\centering
		\includegraphics[width=\textwidth]{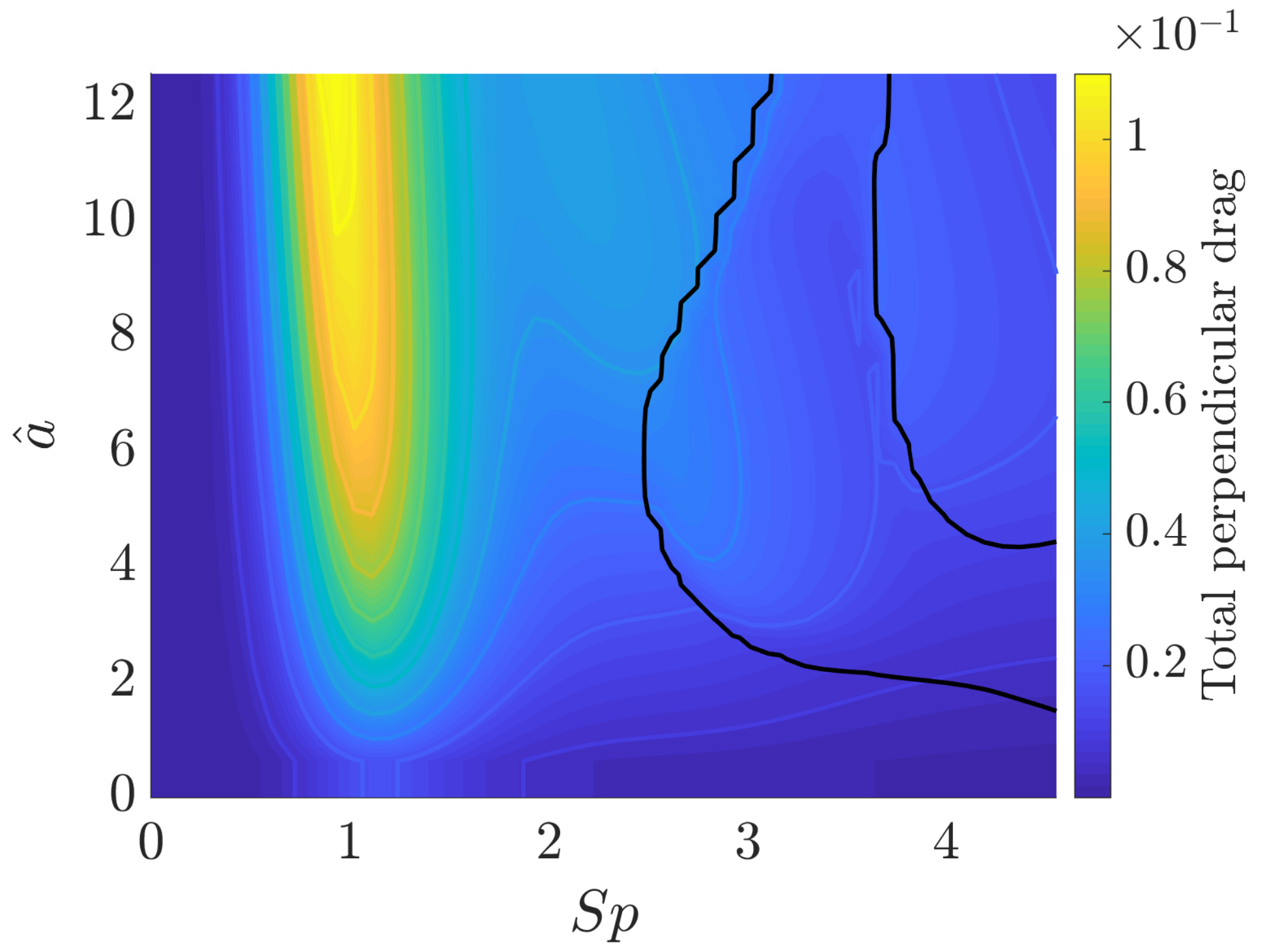}
		\caption{\label{fig:results:oscillatory_drag:drag_y}}
	\end{subfigure}

	\begin{subfigure}[c]{0.45\textwidth}
		\centering
		\includegraphics[width=\textwidth]{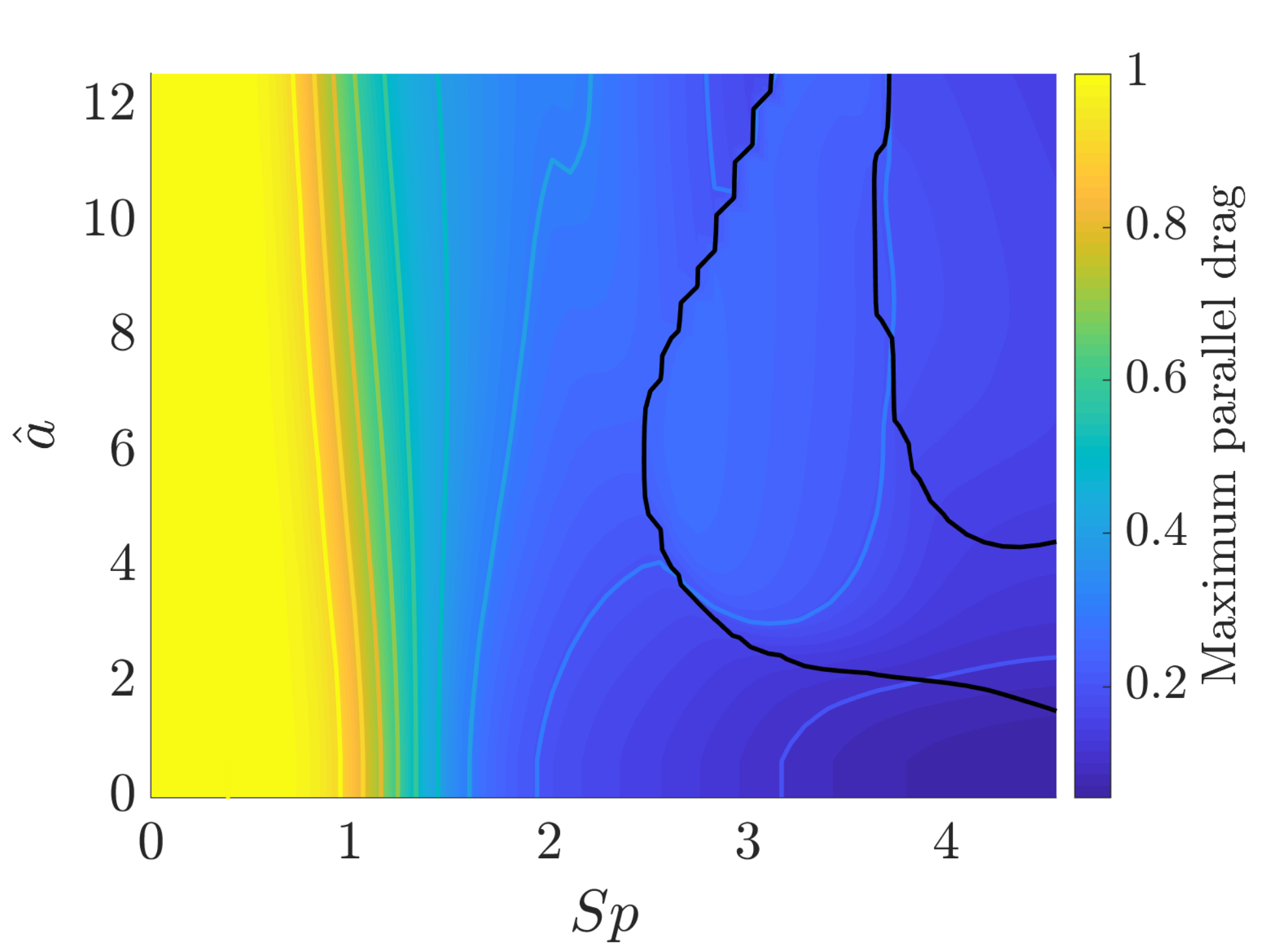}
		\caption{\label{fig:results:oscillatory_drag:max_drag_x}}
	\end{subfigure}
	\begin{subfigure}[c]{0.45\textwidth}
		\centering
		\includegraphics[width=\textwidth]{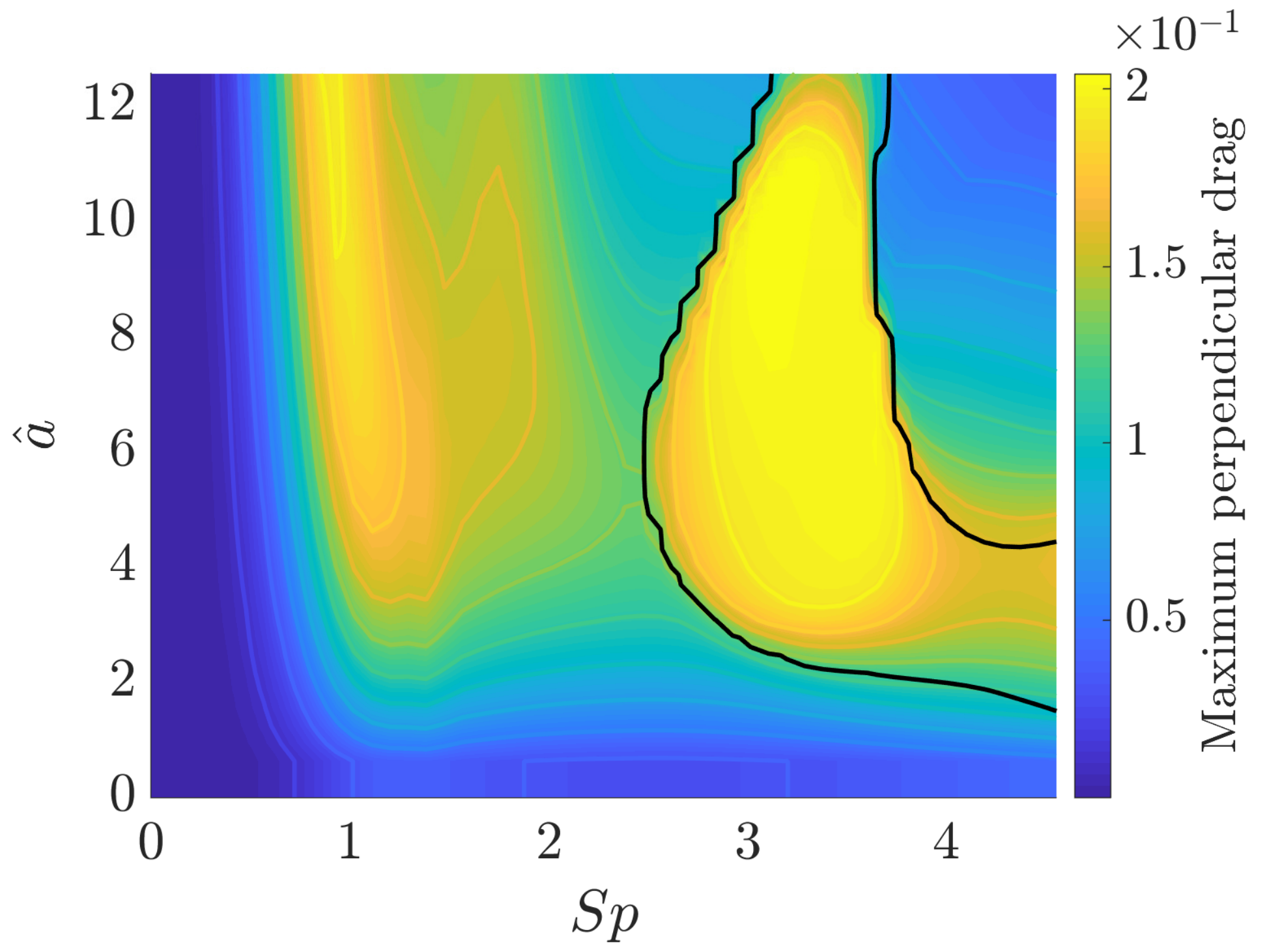}
		\caption{\label{fig:results:oscillatory_drag:max_drag_y}}
	\end{subfigure}
	\caption{Hydrodynamic drag associated with cantilevered filaments in
	oscillating flow over a single period.
	\subref{fig:results:oscillatory_drag:drag_x} Total drag on the filament in
	the direction parallel to the boundary, averaged over a single period. A
	steep increase in total drag is noted to occur upon entering the parameter
	regime of ciliary beating, with a less-prominent local maximum present
	around $\Sp{}=1$.
	\subref{fig:results:oscillatory_drag:drag_y} Total pushing force applied
	on the base of the filament in the direction perpendicular to the boundary
	over a single period of the background flow. A clear maximum can be seen
	around $\Sp{}=1$, corresponding to the symmetric beating of stiff
	filaments, with the high-sperm number border of the ciliary region being
	aligned with a sharp, albeit limited increase in total drag for
	$\hat{a}>6$. \subref{fig:results:oscillatory_drag:max_drag_x} Maximum
	instantaneous total drag on the filament in the direction parallel to the
	boundary. Comparison with \subref{fig:results:oscillatory_drag:drag_x}
	highlights gross total drag cancellation occurring for low \Sp{}, with
	reduced cancellation for the asymmetric ciliary-type beating.
	\subref{fig:results:oscillatory_drag:max_drag_y} Maximum absolute
	instantaneous force applied on the base of the filament in the direction
	perpendicular to the boundary. By comparison with
	\subref{fig:results:oscillatory_drag:drag_y} we observe limited
	cancellation of total drag in this direction in stiff filaments, with
	significant cancellation occurring for ciliary-type beating. Each panel is
	normalised by the maximum instantaneous total parallel drag and rescaled
	by flow amplitude to enable meaningful comparison. Select contours of
	total drag are shown and the region of ciliary-type beating is outlined in
	black, the latter determined only from the kinematic symmetry measure $S$.
	Colour online.
	\label{fig:results:oscillatory_drag}}
\end{figure}

Computation of the total
hydrodynamic drag exerted on the filament by the background oscillatory flow
over a single period reveals a bifurcation structure closely aligned with that
of the morphology of beating. Shown in
\cref{fig:results:oscillatory_drag:drag_x} and rescaled by flow amplitude, the
normalised total drag in the direction parallel to the boundary is seen to be
of constant sign, and minimal for the approximately symmetric beating present
at low sperm numbers, with a region of non-trivial total drag appearing for
$\Sp{}\approx1$ where elastic effects further distance the beating from
reciprocal motion. \cref{fig:results:oscillatory_drag:max_drag_x} demonstrates
that such minimal total drag arises as the result of gross cancellation over
the period of oscillation, consistent with the symmetry of the associated
beating mode. Upon entering the region of parameter space consistent with
ciliary-type beating there is a significant change in the magnitude of the
experienced total drag, intuitively correlated with the emergence of the
distinct forward and backward strokes of ciliary beating. With reference to
the bifurcation diagram of
\cref{fig:results:oscillatory_morphological:bifurcation}, highly-buckled
beating at high values of $\Sp{}$ results in little total drag in the
direction parallel to the boundary, consistent with the truly time-reversible
motion obtained in the limit of
\cref{eq:results:oscillatory_morphological:full_system} as
$\Eh{}\rightarrow\infty$.

Considering similarly the total force applied on the filament base in the
direction normal to the boundary over a single beat period, we again observe
that the total force is of constant sign, corresponding to the filament
pushing towards the boundary. Shown in
\cref{fig:results:oscillatory_drag:drag_y}, the maximal force occurs around
$\Sp{}=1$ for the majority of sampled flow amplitudes, aligning with the local
maximum of total parallel drag noted previously and pertaining to the
non-reciprocal symmetric beating of stiff filaments. Comparison with
\cref{fig:results:oscillatory_drag:max_drag_y} reveals in this case that
significant total drag cancellation occurs for the asymmetric ciliary-type
beating. In contrast, the total drag on stiffer filaments in this direction
sees limited cancellation, exemplifying the significance of beating morphology
in filament drag calculations.

% Given the drag force exerted on the filament by the background, we compute the
% net work done on the filament by the fluid over a single beat period, \note{NOT TRUE, CHANGE}
% naturally having constant negative parity. In
% \cref{fig:results:oscillatory_drag:work} we show the computed work that must
% be done to maintain the background oscillatory flow, related to the work done
% on the filament by a change of sign. Normalised by the maximum attained value
% and rescaled by the square of the flow amplitude, we observed a maximum
% required energy input to the left of the ciliary region, corresponding to
% large-amplitude beating with approximately equal forward and backward strokes.
% Remarkably, crossing into the ciliary region of parameter space results in a
% drastic reduction in required energy input. Thus by this measure we conclude
% that the characteristic and distinct forward and backward strokes of the
% ciliary-type beat are more efficient than other modes of beating observed
% across sperm numbers for a prescribed oscillatory flow, requiring less energy
% input to maintain the periodic beating of the filament and the background
% flow. \note{More detail here, or leave excessive speculation for discussion?
% Open to suggestions for elaborating this section.}

\section{Discussion}
\label{sec:discussion}
%!TEX root=../main.tex
In this work we have examined the accuracy of resistive force theories in
quantifying the mechanics of planar filament motion in a half space via
regularised Stokeslet segments, further exploring the coupled non-local
elastohydrodynamics of cantilevered filaments in an oscillatory flow. We have
seen that the corrections to free-space resistive force theory of
\citet{Katz1975} and \citet{Brenner1962} perform well both in the near and
far-field of planar boundaries when applied to straight filaments, in
agreement with the verification of \citet{Ramia1993} and serving as additional
validation of our methodology.

By considering the motion of curved filaments in planes parallel to a
boundary, in line with our hypothesis we have found that these corrections
capture quantitative filament dynamics to moderate accuracy in all but the
most extreme boundary proximities. Remarkably however, when very close to the
boundary we have nonetheless observed a tightly-distributed ratio of effective
drag coefficients, and indeed tightly-distributed coefficients, suggesting
that a resistive force theory with such coefficients could yield accurate
estimates for hydrodynamic drag in this circumstance. With subjects imaged in
the relative near-field of a coverslip, the similar conclusion of
\citet{Friedrich2010} is thus in part supported by our non-local calculations,
albeit here at greatly-reduced boundary separations, and suggests the future
development of an accurate, empirical resistive force theory in the near-field
of boundaries. This has potential application to the study of a slithering
mode of swimming, as reported to be prevalent in spermatozoa by
\citet{Nosrati2015} though this mode is far from ubiquitous. Analytical
exploration of this result would require consideration of the asymptotic limit
where the separation of singularities from their images is much less than the
lengthscale of the filament radius of curvature, with such detailed
calculations expected to be a subject of significant future study.

We have seen that the agreement between methodologies for parallel-swimming
filaments does not carry through to those moving in a plane perpendicular to
the boundary, with no clear consensus on effective coefficient ratio for
pinned filaments moving perpendicular to boundaries. Thus we conclude that
constant-coefficient resistive force theories cannot be expected to give
accuracy comparable to non-local methods for filaments moving perpendicular to
boundaries. Despite this, here we have reverified the conclusion of
\citet{Curtis2012}, based originally on resistive force theory, though this
appears to rely on serendipitous cancellations on integrating along the
filament. 

In particular, we have noted a potential lack of reliability in local drag
theories when applied to pinned filaments, as evidenced by the loss of
coherence in effective drag coefficient ratio when using artificially-pinned
kinematic data in \cref{sec:results:parallel}, supported further by the
findings of \cref{sec:results:pinned} for tethered spermatozoa. Accordingly,
we have utilised the non-local hydrodynamics of \citet{Cortez2018} to examine
the response of a cantilevered elastic filament to an oscillatory background
flow, with this application highlighting the flexibility in our treatment of
the coupled elastohydrodynamics. Having established convergence to a single
limiting periodic behaviour from a range of initial conditions for fixed
filament and flow parameters, as detailed in the \cref{app:phase}, we have
evidenced in passive flow-driven filaments the existence of a remarkable mode
of beating typically characteristic of actively-beating cilia. The fact that
the filament movement appears qualitatively similar to that of
actively-beating cilia further highlights that pumping fluxes are sensitive to
the details of a ciliary beat with, here, zero total flux of fluid above the
filaments, in contrast to the induced fluxes of ciliary pumping.

The asymmetric ciliary-type mode of beating was found to be accompanied by two
distinct highly-symmetric beating modes. Most prevalent in the
\Spa{} parameter space was that of low-amplitude beating, with the
pinned filament retaining low curvature throughout and generating the maximal
time-averaged total normal force into the boundary amongst beating
morphologies. Highly-buckled modes corresponding to less-rigid filaments
produced minimal time-averaged total drag over each period of the background
flow, with the high-order buckling modes still resolved by our
piecewise-linear formulation of the governing elastohydrodynamics. The
near-symmetry of each of these modes resulted in little time-averaged total
drag in the direction parallel to the boundary. In contrast, entering the
region of parameter space corresponding to ciliary beating is seen to strongly
correlate with a sharp increase in time-averaged total parallel drag,
intuitively the result of kinematic asymmetry arising from the distinct power
and recovery strokes of ciliary beating. Hence, elastohydrodynamically-induced
morphological changes in general have a dramatic effect on filament drag
mechanics. Furthermore, the efficiency of the presented methodology would
facilitate in-depth mechanical studies of primary cilia with regards to
transmitted force and mechanotransduction, noting that the interaction of
bending modulus variations, effective boundary conditions at the ciliary base,
changes in cross section of the cilium, and cilium length are reported to be
under-explored even in the restricted context of renal physiology
\citep{Nag2017a}.

In summary, we have considered in detail the validity of traditional and
corrected resistive force theories for filaments in the presence of a planar
boundary, concluding that, whilst in general these local drag theories may not
be relied upon to perform well against non-local solutions for curved
filaments, for filament motion parallel to a boundary the use of a resistive
force theory may be remarkably accurate with appropriately-chosen
coefficients, and viable over a range of boundary separations. Verified
against previous high-accuracy methods, we have additionally presented an
efficient non-local formulation of the governing elastohydrodynamics in a
half-space, exemplifying its flexibility by application to a cantilevered
filament in oscillatory flow. This study of passive filaments revealed a
surprising asymmetric beating morphology found typically in active cilia,
highlighting a complex relationship between passive elastic fibers,
internally-forced filaments, the flows that they drive and the forces that
they induce.

\appendix
%!TEX root=../main.tex
\section{Influence of flow phase on cantilevered filament dynamics}\label{app:phase}
\begin{figure}
\centering
	\begin{subfigure}[c]{0.32\textwidth}
		\centering
		\includegraphics[width=\textwidth]{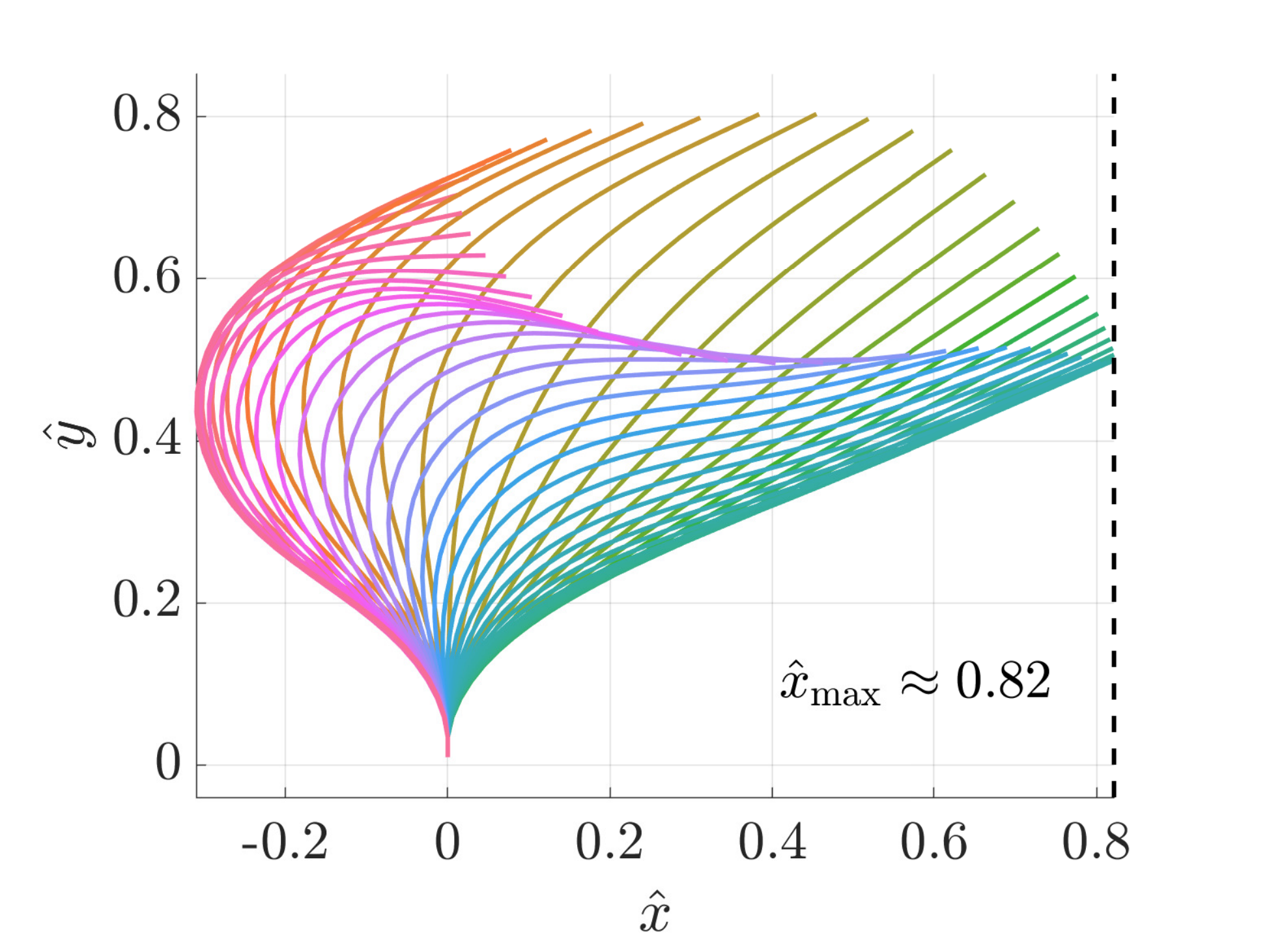}
		\caption{\label{fig:app:phase:def}}
	\end{subfigure}
	\begin{subfigure}[c]{0.32\textwidth}
		\centering
		\includegraphics[width=\textwidth]{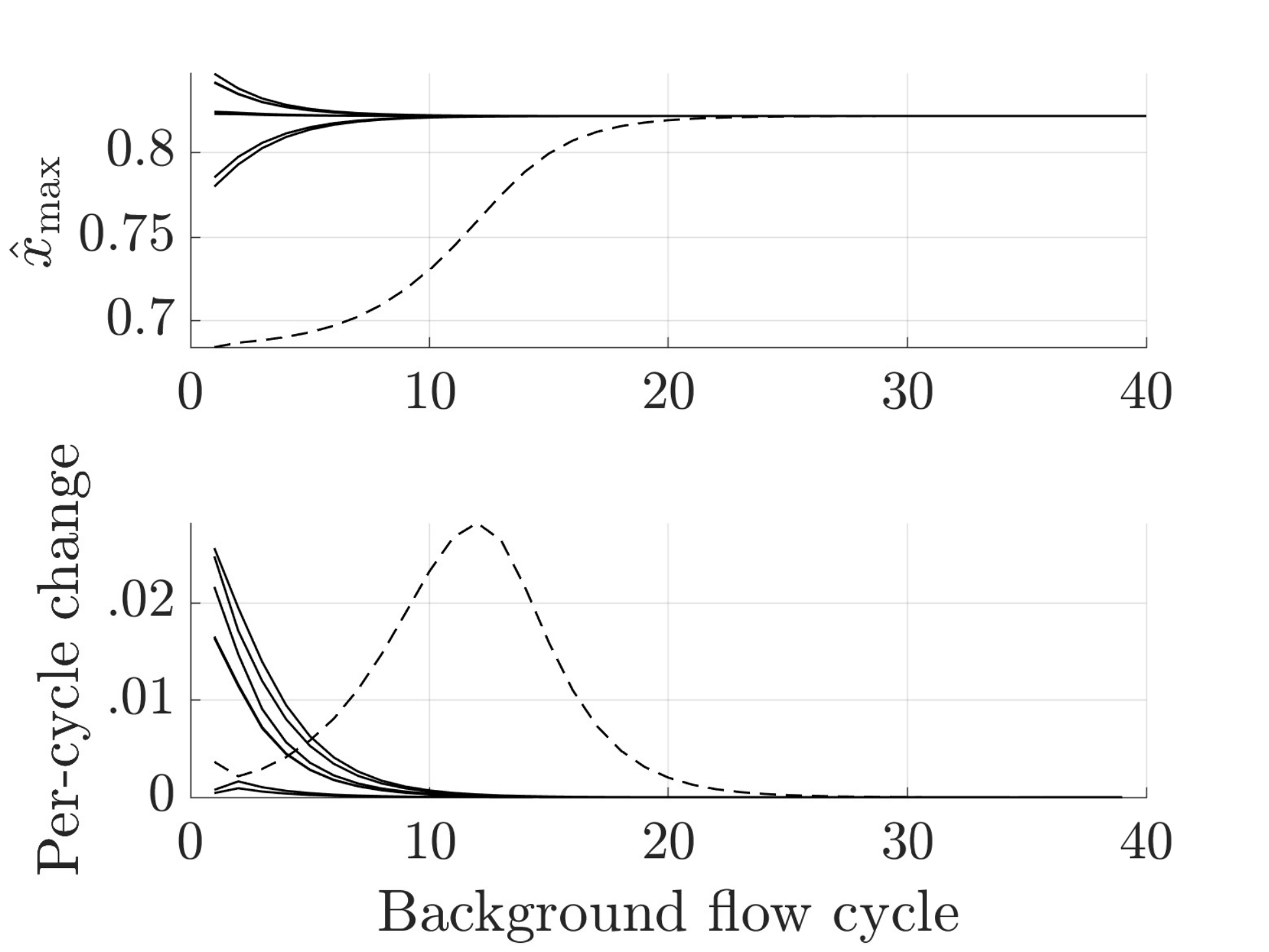}
		\caption{\label{fig:app:phase:convergence}}
	\end{subfigure}
	\begin{subfigure}[c]{0.32\textwidth}
		\centering
		\includegraphics[width=\textwidth]{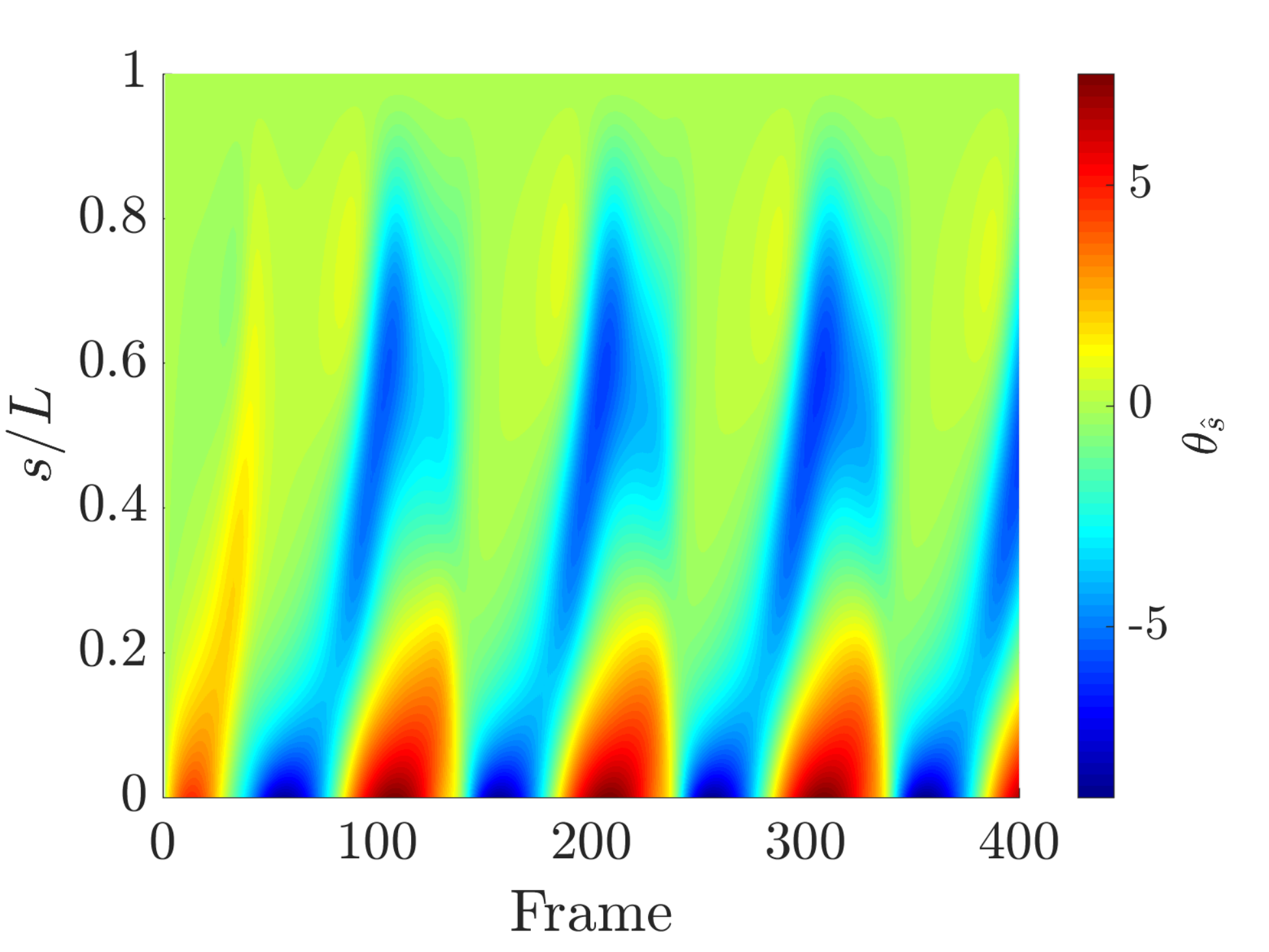}
		\caption{\label{fig:app:phase:curvature}}
	\end{subfigure}
	\caption{Evaluating the sensitivity of long-time behaviour to background
	flow phase. \subref{fig:app:phase:def} An illustration of the measure
	$\hat{x}_{\text{max}}$ used to determine periodicity and convergence of
	filament behaviour, equal to the maximal displacement of the filament tip
	from the centreline over one period of the background flow. Filament
	configurations throughout one period of motion are shown as coloured
	curves. \subref{fig:app:phase:convergence} The evolution of
	$\hat{x}_{\text{max}}$ over 40 periods of the background flow, for various
	choices of initial phase $\hat{t}_0\in[0,\pi]$, here for $\Sp=3.13$,
	$\hat{a}=2\pi$. In the upper panel we see that all choices of $\hat{t}_0$
	lead to convergence to the same behaviour, with even the approximate
	initial symmetry of the $\hat{t}_0=\pi/2$ instance (dashed) collapsing
	onto the common limiting behaviour. A posteriori validation of convergence
	to periodic motion from all phases is provided by consideration of the
	per-cycle change in $\hat{x}_{\text{max}}$, as shown in the lower panel
	and highlighting eventual periodicity. Symmetric counterparts for
	$\hat{t}_0\in[\pi,2\pi]$ have been omitted.
	\subref{fig:app:phase:curvature} Signed curvature of the filament during
	the first periods of the background flow, sampled at 100 frames per
	period, for $\hat{t}_0=3\pi/8$, showing the transition to ciliary beating
	from a straight initial configuration. Colour online.
	\label{fig:app:phase}}
\end{figure}

One might expect the phase $\hat{t}_0$ of the background oscillatory flow of
\cref{sec:results:oscillatory_morphological} to have significant impact on
long-time filament dynamics, the most notable of such effects being the
selection of left-right polarity in resulting behaviours. We examine the
long-time dynamics for a range of phases, flow amplitudes and
elastohydrodynamic numbers, tracking in particular the maximally-attained
displacement of the filament tip from the centreline $\hat{x}=0$ throughout
each period of the background flow, denoted $\hat{x}_{\text{max}}$ and
exemplified by \cref{fig:app:phase:def}. In particular, convergence of
$\hat{x}_{\text{max}}$ to a constant would serve to validate, a posteriori,
the assumption of periodic motion. Sampling $\Sp$ and $\hat{a}$ each from the
range $[0.5,12]$, as expected we observe a duality in solutions inherited from
the left-right symmetry of the problem, and we additionally note the
convergence of all solutions to a single periodic motion, where the period is
aligned with that of the background flow to working precision. This is
demonstrated by \cref{fig:app:phase:convergence}, which shows the evolution of
$\hat{x}_{\text{max}}$ over a number of cycles of the background flow for
$\Sp=3.13$ and $\hat{a}=2\pi$. We see convergence of all solutions to a common
limit cycle of periodic behaviour, with even the approximately-symmetric
transient motion of $\hat{t}_0=\pi/2$ (shown dashed) eventually collapsing
onto the same mode of behaviour. Hence we resolve that exploration of the
$\Sp$--$\hat{a}$ parameter space, holding $\hat{t}_0$ constant, is sufficient
to capture the long-time behaviours of the cantilevered filament system.

\bibliographystyle{jfm}

\end{document}